\documentclass[%
 aip,
 amsmath,amssymb,
 reprint,%
]{revtex4-1}

\usepackage{color}
\usepackage{amsthm}
\usepackage{graphicx}
\usepackage{amsmath}
\usepackage{amssymb}
\usepackage{threeparttable}
\usepackage{graphicx}% Include figure files
\usepackage{dcolumn}% Align table columns on decimal point
\usepackage{bm}% bold math
\usepackage{array}
\usepackage[utf8]{inputenc}
\usepackage[T1]{fontenc}
\usepackage{mathptmx}
\usepackage{etoolbox}

\makeatletter
\def\@email#1#2{%
 \endgroup
 \patchcmd{\titleblock@produce}
  {\frontmatter@RRAPformat}
  {\frontmatter@RRAPformat{\produce@RRAP{*#1\href{mailto:#2}{#2}}}\frontmatter@RRAPformat}
  {}{}
}%
\makeatother

\newcolumntype{C}[1]{>{\centering\arraybackslash}p{#1}}

\begin{document}

%%%%%%%%%%%%%%%%%%%%%%%%%%%%%%%%%%%%%%%%%%%%%%%%%%%%%%%%%%%%
%%%%%%%%%%%%%%%%%%%%%%%%%%%%%%%%%%%%%%%%%%%%%%%%%%%%%%%%%%%%

\preprint{AIP/123-QED}

\title[Bridge equations for transition paths]{Realistic Transition Paths for Large Biomolecular Systems: \\A Langevin Bridge Approach}
\author{Patrice Koehl}
% \email{koehl@cs.ucdavis.edu.}
\affiliation{ 
Department of Computer Science and Genome Center, University of California, Davis, CA 95616, USA.
%\\This line break forced with \textbackslash\textbackslash
}%

\author{Marc Delarue}
% \email{delarue@pasteur.fr}
\affiliation{%
Architecture et Dynamique des Macromol\'{e}cules
Biologiques, UMR 3528 du CNRS, Institut Pasteur, 75015 Paris, France
%\\This line break forced% with \\
}%

\author{Henri Orland}
% \email{henri.orland@ipht.fr}
\affiliation{%
Department of Physics, School of Sciences, Great Bay University, Dongguan, Guangdong 523000, China}
\affiliation
{Universit\'e Paris-Saclay, CNRS, CEA, Institut de Physique Th\'{e}orique, 91191, Gif-sur-Yvette, France
}

\date{\today}% It is always \today, today,

%%%%%%%%%%%%%%%%%%%%%%%%%%%%%%%%%%%%%%%%%%%%%%%%%%%%%%%%%%%%
%%%%%%%%%%%%%%%%%%%%%%%%%%%%%%%%%%%%%%%%%%%%%%%%%%%%%%%%%%%%

\begin{abstract}
We introduce a computational framework for generating realistic transition paths between distinct conformations of large biomolecular systems. 
The method is built on a stochastic integro-differential formulation derived from the Langevin bridge formalism, which constrains molecular trajectories to reach a prescribed final state within a finite time and yields an efficient low-temperature approximation of the exact bridge equation. 
To obtain physically meaningful protein transitions, we couple this formulation to a new coarse-grained potential combining a G\={o}-like term that preserves native backbone geometry with a Rouse-type elastic energy term from polymer physics; we refer to the resulting approach as SIDE.
We evaluate SIDE on several proteins undergoing large-scale conformational changes and compare its performance with established methods such as MinActionPath and EBDIMS. SIDE generates smooth, low-energy trajectories that maintain molecular geometry and frequently recover experimentally supported intermediate states. Although challenges remain for highly complex motions—largely due to the simplified coarse-grained potential—our results demonstrate that SIDE offers a powerful and computationally efficient strategy for modeling biomolecular conformational transitions.
\end{abstract}

%%%%%%%%%%%%%%%%%%%%%%%%%%%%%%%%%%%%%%%%%%%%%%%%%%%%%%%%%%%%
%%%%%%%%%%%%%%%%%%%%%%%%%%%%%%%%%%%%%%%%%%%%%%%%%%%%%%%%%%%%

\maketitle

%%%%%%%%%%%%%%%%%%%%%%%%%%%%%%%%%%%%%%%%%%%%%%%%%%%%%%%%%%%%
%%%%%%%%%%%%%%%%%%%%%%%%%%%%%%%%%%%%%%%%%%%%%%%%%%%%%%%%%%%%
\section{Introduction}
\label{sec:intro}
%%%%%%%%%%%%%%%%%%%%%%%%%%%%%%%%%%%%%%%%%%%%%%%%%%%%%%%%%%%%
%%%%%%%%%%%%%%%%%%%%%%%%%%%%%%%%%%%%%%%%%%%%%%%%%%%%%%%%%%%%

Biomolecules serve as the fundamental workhorses of cells, carrying out nearly all essential biological functions. These functions are governed both by molecular shape -specifically, the three-dimensional (3D) geometries that define biomolecular conformations- and by the inherent change of shape that occurs
as they perform their functions.
Significant progress has been made in characterizing biomolecular structures, aided by a combination of experimental and computational approaches. Experimental methods such as X-ray crystallography, Nuclear Magnetic Resonance (NMR) spectroscopy, and cryo-Electron Microscopy (cryo-EM) have been instrumental in this effort.
A particularly notable achievement in this field is the long-sought ability to accurately predict protein 3D structures directly from their amino acid sequences -a challenge that is now considered largely solved. At the forefront of this breakthrough is AlphaFold \cite{Jumper:2021}, developed by Google DeepMind, whose creators, Demis Hassabis and John Jumper, received the 2024 Nobel Prize in Chemistry for this accomplishment \cite{NobelPrize:2024}. Comparable advances have also been achieved by other methods, including RoseTTAFold, developed in David Baker’s laboratory \cite{Baek:2021, Krishna:2024}, and ESMFold from Meta AI \cite{Lin:2023}.
However, biomolecular function arises not merely from static structures but from structural dynamics—the continuous motions and conformational transitions that occur over a broad range of temporal and spatial scales. Understanding these dynamics remains one of the central challenges in structural molecular biology.
Inspired by the success of deep learning in predicting static protein structures, researchers are now actively exploring other deep learning algorithms aimed at predicting the conformational changes of proteins (see for example Refs \cite{Mollaei:2023, Park:2023, Georgouli:2024, Liu:2025, Pang:2025}). Currently, a major challenge in the development of such models (and in the development of computational studies of protein dynamics in general) lies in the limited experimental training data characterizing dynamics. 
Experimentally, only a few techniques can capture time-resolved structural data, and those that do typically probe restricted temporal windows. 
Computational studies of protein dynamics, especially those aimed at simulating transitions between distinct molecular conformations,
have met with limited success. The core difficulty lies in the fact that such conformational transitions represent rare events relative to the molecule’s internal dynamical time scales. These transitions arise from stochastic fluctuations in molecular structure, driven by thermal energy from the surrounding heat bath, and are infrequent whenever the energy barrier separating two conformations is large compared to the ambient thermal energy, $k_B T$ (where T is the temperature and $k_B$ the Boltzmann constant).

There is actually a lot of information on protein dynamics in the PDB itself.
Indeed, there are many proteins for which multiple conformations have been characterized. 
Those conformations arise because of the formation of a complex (for example, the binding of a ligand) or
changes in the environment (temperature, pH, salt conditions) that influence the stable conformation observed.
What we actually need are techniques that can either capture or predict the changes between those conformations, 
what is referred to as a transition.
%One such techniques rely on understanding the landscape of the structural space captured by those conformations through principal component analysis to bias a Brownian molecular dynamics simulation that steer one conformation to another \cite{Orellana:2016, Orellana:2019}.
Vanden Eijnden and colleagues proposed a general approach to studying those transitions, the Transition Path Theory (TPT) \cite{VandenEijnden:2006, Eijnden:2010, Eijnden:2014}. 
TPT provides a framework for finding the shortest, or most probable transition path between two conformations of a molecule. At zero temperature, the TPT is deemed exact. As such, it has served as a touchstone for the development of many path-finding algorithms.
Some of those were developed for finding the Minimum Energy Path (MEP) on the energy surface for a molecule, such as morphing techniques \cite{Kim:2002, Weiss:2009}, gradient descent methods \cite{Maragakis:2005, Zheng:2007, Tekpinar:2010, PS:2010}, the nudged elastic band method \cite{Jonsson:1998, Henkelman:2000, Sheppard:2008} and the string method \cite{E:2002, Ren:2005, Ren:2007,Vanden:2009, Ren:2013, Maragliano:2014}.
Other algorithms are concerned with either finding the Minimum Free Energy Path (MFEP) on the free energy surface for the molecule,\cite{Maragliano:2006,Pan:2008,Matsunaga:2012,Branduardi:2013}
while others search for paths that minimize a functional, such as Onsager-Machlup functional \cite{Bach:1978, Raja:2025}, 
as implemented in the Minimum Action Path (MAP) methods \cite{Olender:1996, Eastman:2001, Franklin:2007, Faccioli:2006, Vanden:2008, Zhou:2008, Carter:2016, Koehl:2016, Koehl:2024}. \\
In some other methods, the Langevin equation is modified to include a bridge between the two end states and enforces the trajectory to join them \cite{Stuart:2004, Hairer:2007, Orland:2011, Mattingly:2012, Delarue:2017, Koehl:2022}. 
In parallel to those methods,   effort  has  been  dedicated to the development and analysis of Markov State Models
(MSMs) \cite{Chodera:2007, Bowman:2009, Pande:2010}.
MSMs aim at coarse-graining the dynamics of the molecular system via mapping it onto a continuous-time Markov jump process, that is, 
a  process  whose  evolution  involves  jumps  between  discretized  states  representing typical  conformations  of  the original  system.
A similar concept referred to as ``Milestoning" has been proposed  by Elber and coll. performs well \cite{Elber:2004, Elber:2015}.
Note that this list is not meant to be a comprehensive coverage of all existing techniques for finding transition paths, as this is a very active area of research with new techniques proposed every year. 

In preliminary studies by two of the authors  \cite{Franklin:2007, Koehl:2024}, we proposed MinActionPath, a method for computing the MAP between two conformations of a protein on a simplified, two-well free energy surface derived from the ENMs of the two conformations. 
The energy at any conformation is defined as the minimum of the energy functions derived from the two wells. Using this energy functional, the equations of motion corresponding to the MAP are solved analytically in each well, and continuity conditions at the crossing between the two wells define the transition point. MinActionPath has proved useful for example in characterizing the structural reaction path of a tryptophanyl-tRNA synthetase that involves three conformationally distinct states \cite{Lao:2009,Weinreb:2014}. 
There are, however two main drawbacks with MinActionPath that often lead to non-physical paths. First, the ENM potential does not account for steric hindrance, and second, defining the energy surface as the lower envelope of the quadratic wells at the minima leads to problems of smoothness at the crossing between the wells, i.e. to the presence of a cusp in energy at the transition.
 
This paper addresses the shortcomings of MinActionPath for finding a transition path between two protein conformations in two different directions.
First, we propose a different equation of motion that does not assume an energy function that combines quadratic wells at the two conformations.
This equation comes from our recent work on Brownian and Langevin bridge equations  \cite{Orland:2011, Delarue:2017, Koehl:2022}.
Starting from an overdamped Langevin equation modified with a propagator to force a transition to the target conformation,
we have proposed a non-linear stochastic integro-differential equation (which we will refer to as SIDE), whose approximation at low temperature can be solved efficiently and leads to realistic transitions both on simple 2D potentials and for biomolecules \cite{Delarue:2017}.
Second, in contrast to both MinActionPath and our early work on CLD, we define a new coarse-grained potential to represent the protein conformation that does not mix information from the start and target conformations of the protein of interest.
Instead, it considers a Rouse elastic network for each intermediate conformation during the transition, adding collision terms and pseudo-bonded terms to avoid steric clashes and maintain the correct local geometry of the protein.

The paper is organized as follows. In the next section, we describe our equation of motion for finding a transition path between two conformations of a biomolecule and its approximation at low temperature, referred to as SIDE.
We then describe two coarse-grained potentials used to quantify the energy along the transition, a mixed elastic potential described in a preliminary study \cite{Delarue:2017}, and a new mixed G\={o} potential and elastic potential introduced in this study.
The next section shows applications to a few biomolecular systems.
Finally, we conclude the paper with a discussion on extensions of the method, highlighting some of its current limitations.

%%%%%%%%%%%%%%%%%%%%%%%%%%%%%%%%%%%%%%%%%%%%%%%%%%%%%%%%%%%%
%%%%%%%%%%%%%%%%%%%%%%%%%%%%%%%%%%%%%%%%%%%%%%%%%%%%%%%%%%%%

%%%%%%%%%%%%%%%%%%%%%%%%%%%%%%%%%%%%%%%%%%%%%%%%%%%%%%%%%%%%%%%%%%%%%%%%%%%%
%%%%%%%%%%%%%%%%%%%%%%%%%%%%%%%%%%%%%%%%%%%%%%%%%%%%%%%%%%%%%%%%%%%%%%%%%%%%
\section{A bridge equation for generating trajectories}
%%%%%%%%%%%%%%%%%%%%%%%%%%%%%%%%%%%%%%%%%%%%%%%%%%%%%%%%%%%%%%%%%%%%%%%%%%%%
%%%%%%%%%%%%%%%%%%%%%%%%%%%%%%%%%%%%%%%%%%%%%%%%%%%%%%%%%%%%%%%%%%%%%%%%%%%%

The path generation strategy we follow is based on a stochastic integro-differential equation ,
initially referred to as a Langevin bridge (LB) equation \cite{Orland:2011} and later amended as the
the Conditioned Langevin Dynamics (CLD) Equation \cite{Delarue:2017}. 
As it is core to our strategy to derive transition paths, we briefly describe the derivation. 
Full details are available in Refs. \cite{Delarue:2017, Koehl:2022}. 
%We refer to this equation as SIDE.

%%%%%%%%%%%%%%%%%%%%%%%%%%%%%%%%%%%%%%%%%%%%%%%%%%%%%%%%%%%%%%%%%%%%%%%%%%%%
\subsection{The Langevin bridge equation}
%%%%%%%%%%%%%%%%%%%%%%%%%%%%%%%%%%%%%%%%%%%%%%%%%%%%%%%%%%%%%%%%%%%%%%%%%%%%

Consider a system of $N$ particles, each represented
by a position vector ${\bf r}_i \in \mathbb{R}^3, \ i\in \{1,\ldots,N \}$. We are given two conformations for this system,
an initial conformation $I$ and a final conformation $F$.
We want to build a trajectory for the system over a given time $t_f$, such that the system is in state $I$ at $t=0$ and
in state $F$ at $t=t_f$.
We will use ${\bf r}_{i}^{I}$, ${\bf r}_{i}^{F}$, and ${\bf r}_i$ to indicate the position of a particle at time $t=0$, $t=t_f$, and $t$, respectively.
The particles of the system interact through a conservative force derived from a potential $U$. 
The system is evolved using overdamped Langevin dynamics
\begin{equation}
\dot{{\bf r}_i}=-\frac{1}{\gamma}{\nabla}_iU+\bm{\eta_i}(t),
\label{eq:langevin}
\end{equation}
where ${\bf F}_i=-{\nabla}_i U$ is the force acting on particle $i$, $\bm{\eta_i}$
is the Gaussian random force, and $\gamma$ is the friction coefficient.
The potential $U$ may be the sum of a one-body $b$ and two-body $v$ potential
\begin{equation}
U({\bf r}_1,\ldots,{\bf r}_N) = \sum_{i=1}^N b({\bf r}_i) + \frac{1}{2}\sum_{1 \le i\ne j \le N} v({\bf r}_i -{\bf r}_j),
\end{equation}
and of more complicated interaction terms for the realistic atomic description of molecular systems.
The friction coefficient is related to the diffusion constant $D$
and the temperature $T$ through the Einstein relation
\begin{equation}
\gamma=\frac{k_{B}T}{D}=\frac{1}{D\beta},\label{eq:einstein}
\end{equation}
where $\beta=1/k_{B}T$. The friction is usually taken to be independent
of $T$ , so that the diffusion coefficient $D$ is proportional to
the temperature $T$.

The moments of the Gaussian white noise are given by
\begin{eqnarray}
\langle\eta_{i}^k (t) \rangle & = & 0, \nonumber \\
\langle\eta_{i}^k(t)\eta_{j}^l(t')\rangle & = & 2D\delta_{ij} \delta_{kl}\delta(t-t'),\label{eq:noise}
\end{eqnarray}
where the indices $k$ and $l$ denote components of the vector $\bm{\eta}_i(t)$.
As the diffusion constant $D$ is proportional to $T$, the random force $\bm{\eta}_i(t)$ is
of order $\sqrt{T}$.

The probability distribution function $P(\{{\bf r}_i\},t|\{{\bf r}_{i}^{I}\},0)=P(\{{\bf r}_i\},t)$
for the system to be at positions $\{{\bf r}_i\}$ at time $t$ given that
it was at position $\{{\bf{r}}_{i}^{I}\}$ at time 0, satisfies the Fokker-Planck
(FP) equation
\begin{equation}
\frac{\partial P}{\partial t}=D\sum_i {\nabla}_i \left({\nabla}_i P+\beta{\nabla}_i U P\right).\label{eq:FP}
\end{equation}
Among all the paths generated by the Langevin equation (\ref{eq:langevin}),
we are only interested in those that are conditioned to end at a given configuration
$\{{\bf r}_{i}^{F}\}$ at time $t_{f}$.
To this end,
we use the method of Brownian bridges introduced through the Doob transform \cite{Doob:1957}. We denote by $\mathcal{P}(\{{\bf r}_i\},t)$
the probability that the conditioned system is at $\{{\bf r}_i \}$
at time $t$. We have
\begin{equation}
\mathcal{P}(\{{\bf r}_i \},t)=\frac{P(\{{\bf r}_{i}^{F}\},t_{f}|\{{\bf r}_i\},t)P(\{{\bf r}_i \},t|\{{\bf r}_{i}^{I}\},0)}{P(\{{\bf r}_{i}^{F} \},t_{f}|\{{\bf r}_{i}^{I}\},0)}.
\label{eq:proba}
\end{equation}
The probability $P(\{{\bf r}_i\},t|\{{\bf r}_{i}^{I}\},0)$ satisfies equation (\ref{eq:FP})
whereas the function $Q_{1}(\{{\bf r}_i\},t)=P(\{{\bf r}_{i}^{F}\},t_{f}|\{{\bf r}_i\},t)$ above 
satisfies the following reverse or adjoint Fokker-Planck (FP) equation \cite{VanKampen:1992}.

\begin{equation}
\frac{\partial Q_{1}}{\partial t}=-D\sum_i {\nabla}_i^{2}Q_{1}+D\beta \sum_i  {\nabla}_i U .{\nabla}_i Q_{1}.
\label{eq:FPadjoint}
\end{equation}
Using eq.(\ref{eq:FP}) and (\ref{eq:FPadjoint}), one can easily
see that $\mathcal{P}(\{{\bf r}_i\},t)$ satisfies the modified FP equation
\begin{equation}
\frac{\partial\mathcal{P}}{\partial t}=D\sum_i {\nabla}_i\left({\nabla}_i \mathcal{P}+{\nabla}_i\left(\beta U-2\ln Q_{1}\right)P\right)
\label{eq:modifFP}
\end{equation}
and that the positions $\{{\bf r}_i(t)\}$ of the conditioned
system satisfy a modified Langevin equation given by
\begin{equation}
\dot{{\bf r}_i}=-\frac{1}{\gamma}{\nabla}_i U+2D{\nabla}_i \ln Q_{1}+\bm{\eta}(t).
\label{eq:bridge}
\end{equation}
This equation is called a bridge equation \cite{Doob:1957}. The additional force
term $2D{\nabla}_i \ln Q_{1}$ conditions the paths and guarantees that
they will end at $( \{{\bf r}_{i}^{F}\},t_{f})$. We can rewrite $Q_1$ as:
\begin{eqnarray}
Q_{1}(\{{\bf r}_i\},t) &=&e^{-\beta\left(U(\{{\bf r}_i^{F}\})-U(\{{\bf r}_i\})\right)} \langle \{{\bf r}_i^F\} | e^{-(t_f - t) H} | \{{\bf r}_i\} \rangle. \nonumber \\
  \label{QM}
\end{eqnarray}
In equation (\ref{QM}), we have used standard quantum mechanical notation \cite{Feynman:1965} for the matrix element of the evolution operator $e^{-Ht}$ ,  where the Hamiltonian $H$ is given by
\begin{equation}
H=-D\sum_i {\nabla^2_i}+ D\beta^2 W(\{{\bf r}_i\}). 
\label{eq:hamiltonien}
\end{equation}
In equation \ref{eq:hamiltonien}, the potential $W$ is given by
\begin{equation}
W(\{{\bf r}_i\})=\frac{1}{4}\sum_i \left({\nabla}_i U\right)^{2}-\frac{k_{B}T}{2}{\nabla^{2}_i}U.
\label{eq:effective}
\end{equation}
The driving term $Q_{1}(\{{\bf r}_i\},t)$ is a sum over all paths joining $(\{{\bf r}_i\},t)$
to $(\{{\bf r}_{i}^{F}\},t_{f}),$ properly weighted by the so-called Onsager-Machlup
action \cite{Onsager:1953},  $\exp\left(-\frac{1}{4D}\int_{t}^{t_{f}}d\tau\left(\dot{{\bf r}}+\frac{1}{\gamma}{\bf \nabla}U\right)^{2} \right)$.

Defining
\begin{equation}
Q(\{{\bf r}_i\},t)=\langle \{{\bf r}_i^{F}\} | e^{-(t_f - t) H}  | \{{\bf r}_i\} \rangle,
\label{eq:Q}
\end{equation}
the bridge equation (\ref{eq:bridge}) becomes
\begin{equation}
\dot{{\bf r}_i}=2D{\nabla}_i\ln Q+\bm{\eta}_i(t)\label{eq:redbridge}
\end{equation}

In ref. \cite{Koehl:2022}, we have shown that this equation can be exactly recast in the following form
\begin{equation}
\dot{{\bf {r}}_i}=\frac{{\bf r}_{i}^{F}-{\bf r}_i(t)}{t_{f}-t}-\frac{2}{\gamma^{2}}\int_{t}^{t_{f}}d\tau\left(\frac{t_{f}-\tau}{t_{f}-t}\right)\langle{\nabla}_i W({\{\bf r}_i(\tau)\})\rangle_Q+\bm{\eta}_i(t),
\label{eq:star}
\end{equation}
where the bracket $\langle\cdots\rangle _Q$ denotes the average over
all paths joining $(\{{\bf r}_i\},t)$ to $(\{{\bf r}_{i}^{F}\},t_{f}),$ weighted
by the action of equation (\ref{eq:Q})
\begin{eqnarray}
\label{eq:average}
\langle {\nabla}_i W({\bf r}(\tau))\rangle_Q&=&
 \frac{\langle \{{\bf r}_i^{F}\} | e^{-(t_f - \tau) H} \nabla W(\{{\bf r}_i\}) e^{-(\tau-t)H} | {\{\bf r}_i\} \rangle}{\langle \{{\bf r}_i^{F}\} | e^{-(t_f - t) H} | \{{\bf r}_i\} \rangle} \nonumber \\
\end{eqnarray}
and the Gaussian noise is defined by equation (\ref{eq:noise}).
Note
that the first term in the right-hand side of equation (\ref{eq:star}) guarantees that
the constraint ${\bf r}_i(t_{f})={\bf r}_{i}^F$ is satisfied. It is the only term that is singular at time $t_f$, since the integral term does not have any singularity at any time.
In fact, in the case of a free Brownian particle, the potential $W$ vanishes, and we recover the standard equation for free Brownian bridges
\begin{equation}
\dot{{\bf {r}}_i}=\frac{{\bf r}_{i}^{F}-{\bf r}_i(t)}{t_{f}-t}+\bm{\eta}_i(t)\label{eq:free}
\end{equation}

Equation (\ref{eq:star}) is the fundamental equation of motion whose solution defines a transition path between the initial and final conformations. 
This equation is a nonlinear stochastic equation. 
It is Markovian, in the sense that the right-hand side of equation (\ref{eq:star}) depends only on ${\bf r}(t)$. 
However, the presence of the average over all future paths makes it difficult to use. 

%%%%%%%%%%%%%%%%%%%%%%%%%%%%%%%%%%%%%%%%%%%%%%%%%%%%%%%%%%%%%%%%%%%%%%%%%%%%
\subsection{An efficient approximation for weakly dispersed trajectories}
%%%%%%%%%%%%%%%%%%%%%%%%%%%%%%%%%%%%%%%%%%%%%%%%%%%%%%%%%%%%%%%%%%%%%%%%%%%%
We are interested in problems of free energy barrier crossing, which are of importance in many chemical, biochemical, or biological reactions.
In this specific situation of barrier crossing, according to Kramers theory, the total transition time $\tau_{K}$ (waiting + crossing) scales like the exponential of the barrier height
$\exp(\beta \Delta E^{*})$.
In contrast,  the crossing time (Transition Path Time) $\tau_{c}$ scales like the logarithm of the barrier $\log \beta \Delta E^{*}$ \cite{Gopich:2006, Kim:2015, Laleman:2017}.
We have thus $\tau_{c}<<\tau_{K}$.
In this case, the transition trajectories are very weakly diffusive and are thus almost ballistic. 

In reference Ref. (\cite{Delarue:2017}), by using a cumulant expansion, we obtained a simple approximation for the bridge equations in the transition path time regime mentioned above. Using the exact bridge equation (\ref{eq:star}), it is easy to recover this approximate equation. The approximation consists of considering that the paths that connect $\{{\bf r}_i(t)\}$ to $\{{\bf r}_i^{F}(t)\}$ in (\ref{eq:star}) reduce to a straight line connecting the two points. The equation for this straight line is
\begin{equation}
\label{eq:r}
{\bf r}_i(\tau) = \frac{({\bf r}_i^{F}- {\bf r}_i) \tau + t_f {\bf r}_i  - t {\bf r}_i^{F} }  {t_f -t}.
\end{equation}
Consequently, eq.(\ref{eq:star}) becomes
\begin{equation}
\label{eq:approx1}
\dot{{\bf {r}}_i}=\frac{{\bf r}_{i}^{F}-{\bf r}_i(t)}{t_{f}-t}-\frac{2}{\gamma^{2}}\int_{t}^{t_{f}}d\tau\left(\frac{t_{f}-\tau}{t_{f}-t}\right){\nabla}_i W({\{\bf r}_i(\tau)\})+\bm{\eta}_i(t)
\end{equation}
where ${\bf r}_i(\tau)$ is given by eq.(\ref{eq:r}).

Making the change of variable
\begin{equation}
u=\frac{\tau-t}{t_f -t},
\end{equation} 
we have
\begin{equation}
\mathbf{r}_i(u) = u \mathbf{r}_i^{F} + (1-u) \mathbf{r}_i.
\end{equation}
Equation (\ref{eq:approx1})
becomes
\begin{equation}
\label{approx4}
\dot{{\bf {r}}_i}=\frac{{\bf r}_{i}^{F}-{\bf r}_i(t)}{t_{f}-t}-\frac{2}{\gamma^{2}}(t_f-t)\int_{0}^{1}du (1-u){\nabla}_i W(\{{\bf r}_i(u)\}) +{\bf \eta}_i(t).
\end{equation}

This equation is an integro-differential stochastic Markov equation,  as the variable $\dot {\bf r}_i(t)$ depends only on the stochastic variable $\mathbf{r}_i(t)$ at time $t$. One can generate many independent trajectories by integrating this equation with different noise histories ${\bf \eta}_i(t)$.
It is the basis of the transition path method presented in this paper.
%

%%%%%%%%%%%%%%%%%%%%%%%%%%%%%%%%%%%%%%%%%%%%%%%%%%%%%%%%%%%%%%%%%%%%%%%%%%%%
\subsection{Solving the bridge equation}
%%%%%%%%%%%%%%%%%%%%%%%%%%%%%%%%%%%%%%%%%%%%%%%%%%%%%%%%%%%%%%%%%%%%%%%%%%%%

We found that a simple method to solve the equation is to use a Euler-Maruyama \cite{Kloeden:1992} discretization scheme for the equation, dividing the time $t_{f}$ in $N$ intervals of size $dt$, so that $t_{f}=Ndt$.  
In this scheme, the state of the molecule $\{ {\bf r}(k,i) \}$ at time $(k+1)dt$ is computed from the state at time $kdt$:
\begin{eqnarray}
{\bf r}(k+1,i)=&&{\bf r}(k,i)dt +\frac{{\bf r}^{F}(i)-{\bf r}(k,i)}{t_{f}-kdt}dt -\frac{2}{\gamma^{2}}(t_{f}-kdt) I(k,i) \nonumber \\
&& +\sqrt{2Ddt}\bm{\xi}(k,i),
\label{eq:discret}
\end{eqnarray}
where $\bm{\xi}(k,i)$  are normalized Gaussian variables:
\begin{equation}
\langle\xi^{(a)}(k,i)\rangle = 0; \quad \quad \langle (\xi^{(a)}(k,i))^2 \rangle = 1.
\end{equation}
The integral $I(k,i)$ is computed numerically over $M$ steps within the interval $[0,1]$,
\begin{eqnarray*}
I(k,i) = \frac{1}{M}  \sum_{l=0}^{M-1} (1-\frac{l}{M}){\nabla}_i W(\{{\bf r}(l,i)\}),
\end{eqnarray*}
where
\begin{eqnarray*} 
{\bf r}(l,i) = \frac{l}{M} {\bf r}^F(i) + \left(1 -\frac{l}{M}\right)  {\bf r}(k,i).
\end{eqnarray*}

%%%%%%%%%%%%%%%%%%%%%%%%%%%%%%%%%%%%%%%%%%%%%%%%%%%%%%%%%%%%%%%%%%%%%%%%%%%%
%%%%%%%%%%%%%%%%%%%%%%%%%%%%%%%%%%%%%%%%%%%%%%%%%%%%%%%%%%%%%%%%%%%%%%%%%%%%
\section{Coarse-grained potentials for protein transition paths}
%%%%%%%%%%%%%%%%%%%%%%%%%%%%%%%%%%%%%%%%%%%%%%%%%%%%%%%%%%%%%%%%%%%%%%%%%%%%
%%%%%%%%%%%%%%%%%%%%%%%%%%%%%%%%%%%%%%%%%%%%%%%%%%%%%%%%%%%%%%%%%%%%%%%%%%%%
The previous section introduced an integro-differential equation for computing the trajectory between 2 conformations of a biomolecule.
This equation is valid for any potential $U$ that describes the stability and geometry of the biomolecule considered.
However, this potential is essential for generating realistic trajectories.
We describe two such potentials, both coarse-grained, i.e., based on a simplified representation that includes only one atom per residue, the C$_{alpha}$ for a protein.

%%%%%%%%%%%%%%%%%%%%%%%%%%%%%%%%%%%%%%%%%%%%%%%%%%%%%%%%%%%%%%%%%%%%%%%%%%%%
\subsection{A mixing potential for transition paths for proteins driven by Langevin bridge: the CLD framework}
%%%%%%%%%%%%%%%%%%%%%%%%%%%%%%%%%%%%%%%%%%%%%%%%%%%%%%%%%%%%%%%%%%%%%%%%%%%%

We are concerned with the study of conformational changes between two states of a protein.
We use a coarse-grained representation of the protein structure, namely, we only consider one atom per residue, its $C_{\alpha}$.
In our previous study for deriving protein transition paths, we defined an energy function that is the combination of a mixed elastic model and a collision term
\begin{eqnarray}
U =U_{Mix-ENM} + U_{collision}.  
\end{eqnarray}
In this equation the mixed elastic potential is defined as
\begin{eqnarray}
U_{Mix-ENM}=-\frac{1}{\beta_m} log(e^{-\beta_m U_I} + e^{-\beta_m U_F}),
\end{eqnarray}
where $\beta_m$ is the inverse of the mixing Temperature $T_m$, $U_I$ and $U_F$ are the elastic potential centered on conformation $I$ (initial) and  conformation $F$ (final), respectively \cite{Zheng:2007}. 
The two elastic potentials follow the original definition of Tirion \cite{Tirion:1996}:
\begin{eqnarray}
U_I&=& \sum_{ij} k_{ij} C_{ij} (r_{ij}-r_{ij}^{I})^{2} \nonumber \\
U_F&=& \sum_{ij} k_{ij} C_{ij} (r_{ij}-r_{ij}^{F})^{2} + \Delta U_0
\label{eqn:tir}
\end{eqnarray}
where $C_{ij}$ is a contact matrix that is set to $1$ if $d_{ij} < R_c$ and $0$ otherwise and $k_{ij}$ is its associated elastic constant.  
If a pair (i,j) is present in both forms, we take the same elastic constant $k_{ij}$ for both (see below).
$\Delta U_0$ is the (preset) energy difference between the two states and $r_{ij}^I$ and $r_{ij}^{F}$ their interatomic distances at rest in conformations A and B, respectively. 
The elastic constants $k_{ij}$ are modulated by the difference in the resting $r_{ij}$ distances of the two states.
\begin{eqnarray}
k_{ij}=\min\left(\frac{\epsilon_k}{(r_{ij}^{I}-r_{ij}^{F})^{2}},k_{max}\right)
\label{eqn:k}
\end{eqnarray}

The collision energy term is taken as the repulsive part of a Lennard-Jones potential,
\begin{eqnarray}
U_{collision}=\epsilon \sum_{i,j} \left(\frac{\sigma}{d_{ij}}\right)^{12}
\end{eqnarray}

The potential $W$ is given by,
\begin{eqnarray}
W= \frac{1}{4}  \nabla U ^2 - 0.5 k_B T \Delta U
\end{eqnarray}
Note that $W$ is not a linear combination of an effective energy associated with the elastic term and an effective energy for the collision term, du to the presence of the norm squared of $\nabla U$: there is a cross-term between the two types of potential.
All the algebra needed to implement this energy $U$ and its associated effective energy was described in detail in an Appendix in \cite{Koehl:2016}.

It is important to notice that $U$, and consequently $W$ depends on many parameters.
Those include the mixing temperature $T_m$, the cutoff $R_c$ for the elastic networks, $\epsilon_k$ and $k_{max}$ for the mixing elastic constants, $\epsilon$ and $\sigma$ for the collision term, and $\Delta U_0$ defining the energy difference between states $A$ and $B$.
In addition, this energy function is heavily conditioned by the conformations $A$ and $B$ through their elastic potentials.
This leads to some ambiguities for some terms. For example, $U_I$ and $U_F$ may include very different numbers of pairs of atom, leading to possible biases in the mixing energy. 
We have seen that parameterizing this potential is not easy, leading to cases in which we could not build a transition path based on this energy (the ATPase and RNase cases mentioned in Ref. \cite{Delarue:2017}).
This lead us to propose a different type of potential, presented below.

%%%%%%%%%%%%%%%%%%%%%%%%%%%%%%%%%%%%%%%%%%%%%%%%%%%%%%%%%%%%%%%%%%%%%%%%%%%%
\subsection{A G\={o} like potential for transition paths for proteins driven by Langevin bridge: the SIDE framework}
\begin{figure}[htb]
\centering
\includegraphics[width=0.35\textwidth]{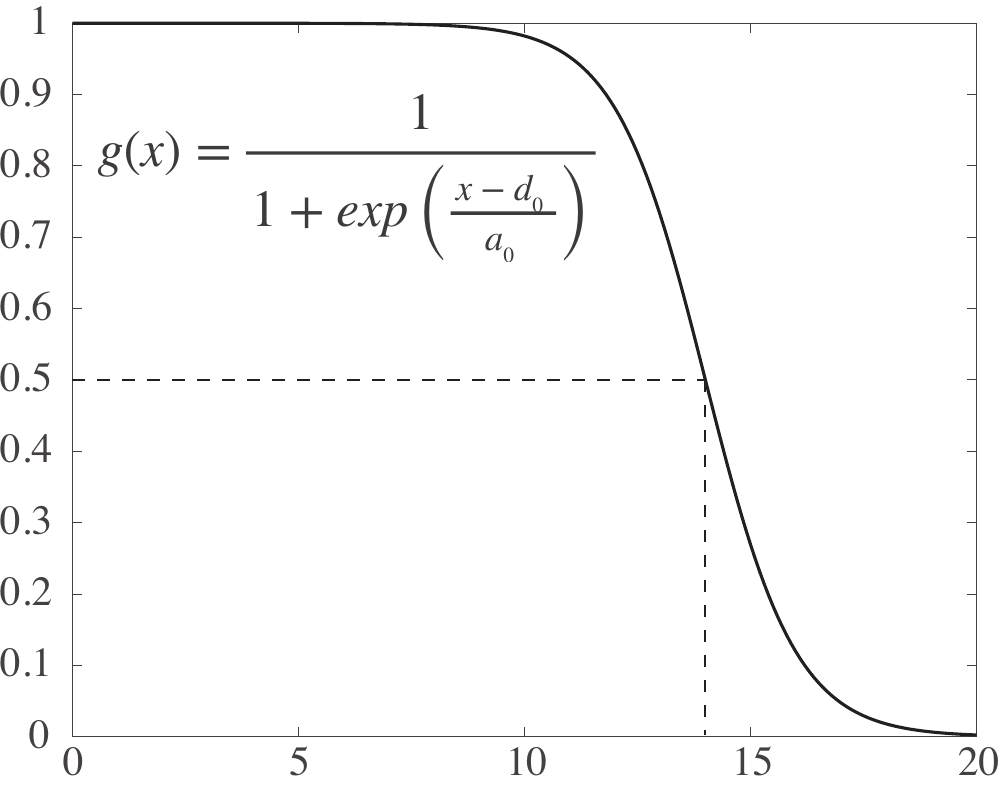}
\caption{\textbf{The Fermi like function $g(x)$} (shown for $d_0=14$ and $a_0=1$).}
\label{fig:fermi}
\end{figure}

%%%%%%%%%%%%%%%%%%%%%%%%%%%%%%%%%%%%%%%%%%%%%%%%%%%%%%%%%%%%%%%%%%%%%%%%%%%%
When designing a new potential for building transition paths between two conformations of a protein, our specifications were double: minimize the number of parameters, and reduce the dependencies to the initial and final conformations. We chose a G\={o}-like potential
as it verifies these two requirements.
The G\={o}-like potential only considers the C$_\alpha$ of all residues in the molecule of interest. In addition to the notations already introduced above, we consider also $\theta_i$, the virtual bond angle formed by the C$_{\alpha}$s of the consecutive residues $i$, $i+1$, and $i+2$.
Our G\={o}-like potential at a conformation $\mathbf{X}$ is defined as (see \cite{Clementi:2000} as well as \cite{Na:2014c} for equivalent potentials):

\begin{eqnarray}
U(\mathbf{X}) &=&U_{b}(\mathbf{X}) + U_{\theta}(\mathbf{X}) + U_{vdW}(\mathbf{X}) + V_{el}(\mathbf{X}) \nonumber \\
\label{eqn:go}
\end{eqnarray}
with:
\begin{eqnarray*}
U_{b}(\mathbf{X}) &=&\frac{100 \epsilon_G }{2} \sum_{i=1}^{N-1}  (r_{i,i+1} - r_{i,i+1}^I )^2,
\end{eqnarray*}
\begin{eqnarray*}
U_{\theta}(\mathbf{X}) &=&  \frac{40 \epsilon_G}{2} \sum_{i=1}^{N-2}   (\theta_i - \theta_i^I )^2,
\end{eqnarray*}
\begin{eqnarray*}
U_{vdW}(\mathbf{X})&=&  \epsilon_{G} \sum_{(i<j-3)}  \left [ \left( \frac{r_{ij}^I}{r_{ij}} \right)^{12} -  \left( \frac{r_{ij}^I}{r_{ij}} \right)^{6} \right],
\end{eqnarray*}
\begin{eqnarray*}
U_{el}(\mathbf{X}) &=&  \frac{\epsilon_G}{2 Np} \sum_{i}\sum_{j}  g(r_{ij}) r_{ij}^2.
\end{eqnarray*}
$I$ stands for the initial conformation, $N$ is the total number of C$_\alpha$ in the protein, $Np$ is the number of pairs considered, $\epsilon_G$ is a constant, and $g(r_{ij})$ is a Fermi-like function that provides a smooth cutoff:
\begin{eqnarray}
g(r) = \frac{1}{1 + \exp{\left(\frac{r-d_0}{a_0}\right)}}
\label{eqn:fermig}
\end{eqnarray}
with $d_0$ and $a_0$ constants (see Figure \ref{fig:fermi}). 

The potential defined in equation \ref{eqn:go} contains three types of terms:
\begin{itemize}
\item [a)] \emph{Bonded interactions}. Both $U_b$ (a bond potential associated with consecutive C$_\alpha$s) and $U_{\theta}$ (an angle potential associated with 3 consecutive C$_\alpha$s) are local potentials meant to maintain the geometry of the protein chain. Two consecutive C$_\alpha$s are at $3.8$ \AA\  or $2.9$ \AA\  apart, depending on the peptide bond being trans or cis, respectively. We capture this information from the initial conformation. The angles associated with  3 consecutive C$_\alpha$s is loosely conserved in protein, but we keep it close to their values in the initial conformation.
\item [b)]  \emph{A vdW term}. $U_{vdW}$ is a 12-6 Lennard-Jones potential set to avoid steric clashes during the transition.
\item [c)] \emph{An elastic potential}. The elastic model is different from a standard Tirion-like elastic model in that it does not include a reference structure. It is replaced by the Rouse Model, a coarse-grained model well studied in polymer physics with springs between beads \cite{Rouse:1953, deGennes:1979}. 
\end{itemize}

The vdW term is limited to pairs of atoms that are within a cutoff distance $R_c$ from each other.
The elastic potential handles the cutoff differently, introducing the Fermi-like function $g(r)$ defined in Equation \ref{eqn:fermig}.
The reason is simple: the vdW potential naturally approaches 0 at large distances because of the 12 and 6 powers of the Lennard Jones potential. In contrast, the elastic potential is proportional to $r^2$, i.e., it becomes large for large distances. 
The presence of the Fermi function imposes a potential of 0 close to $d_0+a_0$. 
There is another reason, however, for the Fermi potential. 
In its absence, the elastic potential would be proportional to the radius of gyration, squared. As such, optimizing the elastic potential would lead to proteins that are as compact os possible, which is not the desired outcome in most cases.

As written, the G\={o} like potential we consider has 4 parameters: the cutoff $R_c$ and its equivalent $d_0$ for the Fermi-like function, the width of the latter, $a_0$, as well as the factor $\epsilon$ common to all terms.
The accepted range of values for $R_c$ in the context of a C$_{\alpha}$-only elastic model is 12 to 14 $\AA$. 
In all experiments described below, we use $R_c=d_0=14$, $a=1$, and $\epsilon = 1$.

All the algebra  needed to implement the G\={o}-like energy $U$ and its associated effective energy ia described in detail in Appendices
\ref{sec:A1} (pairwise interactions), \ref{sec:A2} (angular term), and \ref{sec:A3} (effective energy).

%%%%%%%%%%%%%%%%%%%%%%%%%%%%%%%%%%%%%%%%%%%%%%%%%%%%%%%%%%%%%%%%%%%%%%%%%%%%
\subsection{Principal component analyses of an ensemble of protein structures}
%%%%%%%%%%%%%%%%%%%%%%%%%%%%%%%%%%%%%%%%%%%%%%%%%%%%%%%%%%%%%%%%%%%%%%%%%%%%
A transition path between two structures of a protein is a sampling of the ensemble of conformations that are accessible to the protein along that path. 
It is possible to extract information from that ensemble using statistical techniques.
In particular, principal component analysis (PCA) is well suited for that task as it enables describing the ensemble through a decomposition process that filters observed motions from the most significant to the least significant scales (see for example Ref. \cite{Abdi:2010}).

PCA has been widely used to describe the essential motions of proteins from MD simulations \cite{Amadei:1993, Berendsen:2000}. 
It starts from a coordinate matrix, $X$ of size $3N_a \times N_s$ where $N_a$ is the number of atoms considered, and $N_s$ the number of conformations in the ensemble (in our case, the number of images along the trajectory).  The first step is to remove overall translations and rotations by aligning each conformation to a reference structure (usually the first structure).
Then a new matrix $X'$ is built through centering each row of $X$. A covariance matrix $C=XX'$ is constructed and diagonalized:
\begin{eqnarray*}
C = X'X'^T = E V E^T
\end{eqnarray*}
The column $\mathbf{E}_i$ of $E$ correspond to the $i-th$ principal components (PC), with contribution $\frac{v_i}{\sum_k v_k}$ (the variance of the principal component.
Usually the top $M$ (i.e. with the largest variance) two PCs are chosen to capture a low-dimensional representation of the structure space associated with the $N_s$ conformation in the ensemble.

Note that the matrix $C$ has size $3N_a \times 3N_a$. For a large molecular system, it becomes too large to fit on standard computer memory. However, there
is no need to compute $C$ explicitly. Indeed, we are only interested in a few of the largest eigenvalues of $C$ that can be computed using an iterative method
such as the power method with explicit deflation (see Ref. \cite{Saad:2011}). The idea is the following: after finding the largest eigen pair $(v_1, \mathbf{E}_1)$ of $C$, deflate the matrix:
\begin{eqnarray*}
C_1 = C - v_1 \mathbf{E}_1 \mathbf{E}_1^T.
\end{eqnarray*}
Then apply the power method to $C_1$ to find $v_2$, and repeat until the $M$ top eigen pairs of $C$ have been found.
The power method requires computing the products of $C$ with vectors $V$ (that ultimately converge to an eigenvector of $C$). This is done by computing $X'X'^T V$, removing the need to compute $C$ explicitly. Finally, the variance $\sigma_i$ by an eigenvalue $v_i$ is:
\begin{eqnarray*}
\sigma_i = \frac{v_i}{\sum_k v_k} = \frac{v_i}{\text{tr}(C)} =  \frac{v_i}{ ||X'||_F^2}
\end{eqnarray*}
where $\text{tr}(C)$ is the trace of $C$, and $||X'||_F$ is the Frobenius norm of $X'$.

The PCA procedure described above can be used to compare trajectories as follows.
First, we collect all conformations along all the trajectories to analyze, defining an ensemble.
A coordinate matrix $X$ is then built from this ensemble, and analyzed using the PCA procedure described.
The two main principal components, PC1 and PC2, with their eigenvectors $\mathbf{E}_1$ and $\mathbf{E}_2$ define a space for the ensemble with reduced dimensionality.
The coordinates $c(\mathbf{X})_i$ with $i\in [1,2]$ , of any conformation $\mathbf{X}$ in that space are then computed as 
\begin{eqnarray*}
c(\mathbf{X})_i = < (\mathbf{X}-\mathbf{X}_0) , \mathbf{E}_i >
\end{eqnarray*}

%%%%%%%%%%%%%%%%%%%%%%%%%%%%%%%%%%%%%%%%%%%%%%%%%%%%%%%%%%%%%%%%%%%%%%%%%%%
\section{Results and discussion: Structural transitions for large molecular systems}
\label{sec:results}
%%%%%%%%%%%%%%%%%%%%%%%%%%%%%%%%%%%%%%%%%%%%%%%%%%%%%%%%%%%%%%%%%%%%%%%%%%%

 \begin{figure}[htb]
\centering
\includegraphics[width=0.45\textwidth]{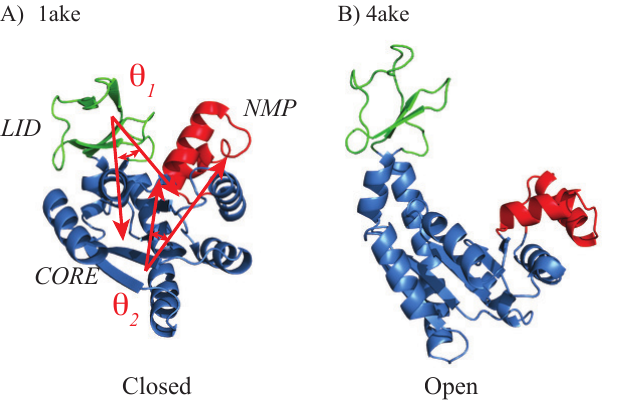}
\caption{\textbf{ Conformational transitions in Adenylate kinase (AKE}. (A) the open state (PDB code: 4AKE); and (B) the closed state  (PDB code: 1AKE). AKE consists of three well-defined domains, the rigid CORE (blue, residues 1–29, 60–121, and 160–214) the nucleotide triphosphate binding domain, LID (green, residues 122–159), and the nucleotide monophosphate binding domain NMP (red, residues 30–59). 
The angle LID-CORE $\theta_1$ is formed by the centers of mass of the backbone of residues of LID ( residues 123–155 (LID), hinge (residues
161–165), and CORE (residues 1–8, 79–85, 104–110, and 190–198), whereas the angle NMP-CORE $\theta_2$ is formed by the centers of mass of the backbone of residues of NMP (residues 50–59), CORE ( residues 1–8, 79–85, 104–110, and 190–198), and hinge (residues 161–165).}
\label{fig:ake}
\end{figure}   

\begin{figure*}
\centering
\includegraphics[width=0.9\textwidth]{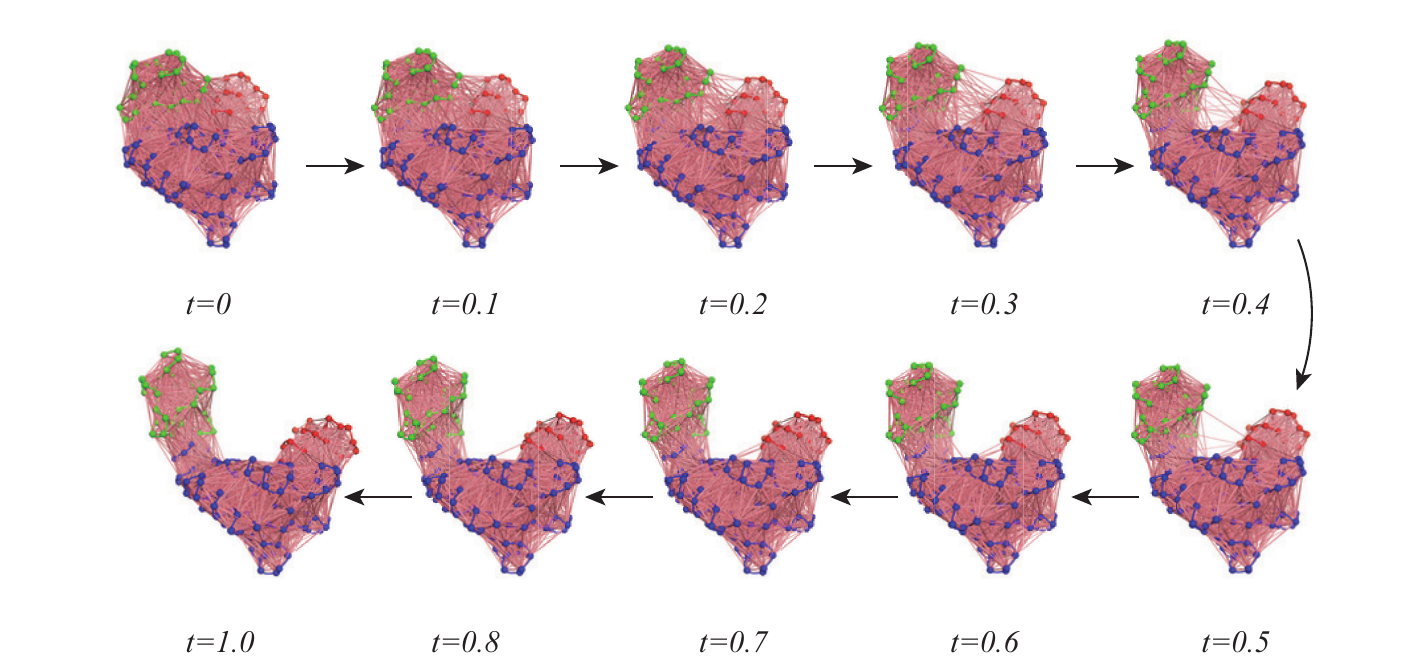}
\caption{\textbf{ SIDE trajectory between the open state and closed state of AKE. } The protein is represented as a string of beads, corresponding to the C$_{\alpha}$ atoms along its main chain, and colored based on its domain definition (see Figure \ref{fig:ake} for details). The elastic network at each pose is illustrated with edges colored in salmon pink. The timing for each pose is given as a fraction of the total time given for the construction of the path.}
\label{fig:ake_traj}
\end{figure*}   

We test three different strategies for generating a trajectory between two conformations of a protein.

The first strategy relies on solving the Langevin bridge equation discussed in the method section. 
We apply it with two different potentials,  a mixed elastic potential (the CLD method) and the new potential introduced in this study
(the SIDE method).

The second strategy, which we will refer as MAP, assumes a harmonic potential at
the end states and solves the action minimization problem by integrating the equations of motion at each end state and finding the crossing point of these two solutions, referred to as a transition state. 
This method, initially referred to as MinactionPath \cite{Franklin:2007}, is available through a web server \cite{Koehl:2024}. We used instead
a standalone version that is equivalent to the web version.

The third strategy, EBDIMS \cite{Orellana:2016}, starts with a stochastic simulation following a Langevin equation,
\begin{eqnarray*}
m_i \ddot{{\bf r}_i} = -\nabla_i U - \gamma \dot{{\bf r}_i} + \bm{\eta}(t),
\end{eqnarray*}
where $U$ is an elastic potential,
\begin{eqnarray*}
U = \frac{1}{2} \sum_{i} \sum_{j} (r_{ij}-r_{ij})^0)^2, 
\end{eqnarray*}
 namely, a Tirion potential \cite{Tirion:1996} (see equation \ref{eqn:tir} for details), and $\bm{\eta}$ is a Gaussian random force.
 The Langevin dynamics starts at one of the two conformations of the protein under study. 
 It is biased in the direction of the other conformation by computing every $k$ steps a progress variable that compares the interatomic distances in the current conformation with the same distances in the target conformation, and accepting a move only if it reduces this progress variable \cite{Orellana:2016}. 
 EBDIMS is available through a web server \cite{Orellana:2019}, and as a standalone program. We used the latter.
 
 The four different programs, CLD, SIDE, PATH, and EBDIMS were run with the following parameters.
 \begin{itemize}
 \item [a)] For CLD, we used the parameters described in Ref. \cite{Delarue:2017}. For its mixed-ENM potential, we set $R_c=11.5 \AA$, and a mixing Temperature $T_m= 1500 T$, where T is taken to be $T=5$.\\
For the Langevin dynamics equation, we set $dt=0.001$ and $\gamma=1$.
\item [b)] For the SIDE potential, we set $\epsilon = 1$ and $R_c=14$. The Fermi function is set with $a_0=1 \AA$ and $d_0 = R_c = 14 \AA$.
For the Langevin dynamics, we set $dt=0.001$, $\gamma = 1$, and $T = 1$.
\item[c)] MinActionPath has two options for defining elastic network, one based on a cutoff distance and the second that derives the elastic network from the Delaunay complex over the C$_{\alpha}$ of the protein. We used the former, setting $R_c=14 \AA$  to match with the network defined with the three other methods. Similarly, MinActionPath includes a Tirion potential and a G\={o} potential; we use the former. This specific version of MinActionPath is referred to as MAP in the following. The trajectories are computed with $t_f = 50$ and $T=1$.
\item[d)] EBDIMS has two main accessible parameters, a cutoff value, and the number of unbiased step of Langevin dynamics, $k$. We have set the cutoff to $6 \AA$ and $k$ to 1, as recommended by the EBDIMS web server.
\end{itemize}
 
%%%%%%%%%%%%%%%%%%%%%%%%%%%%%%%%%%%%%%%%%%%%%%%%%%%%%%%%%%%%%%%%%%%%%%%%%%%%%
\subsection{An illustrative example: Adenylate kinase (AKE) }
%%%%%%%%%%%%%%%%%%%%%%%%%%%%%%%%%%%%%%%%%%%%%%%%%%%%%%%%%%%%%%%%%%%%%%%%%%%%%

 \begin{figure*}[htb]
\centering
\includegraphics[width=0.6\textwidth]{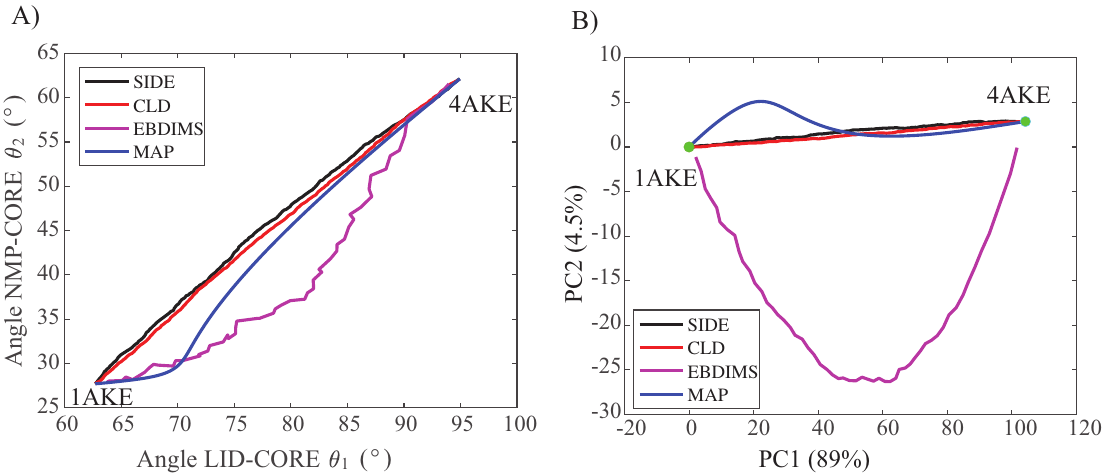}
\caption{\textbf{ Four different trajectories between the open state and closed state of AKE. } We computed trajectories between the open state (1ake) and the closed state (4ake) using SIDE (black), CLD (red), EBDIMS (magenta) and MAP (blue). (A) The angle between the AMP binding domain (NMP) and the CORE is plotted against the angle between the LID and the CORE. (B) Projections of the 4 trajectories along the 2 principal components of their conformational spaces.}
\label{fig:ake_comp}
\end{figure*}   

Adenylate kinase, or AKE in short, is an ubiquitous enzyme that catalyzes the reversible phosphoryl transfer of AMP and ATP into two ADP.
AKE comprises of three main domains, the ATP-binding domain also referred to as LID, the AMP binding domain (NMP)
and the remainder, referred to as the CORE domain (see Figure \ref{fig:ake}).
AKE undergoes a large-scale conformational change between open and closed states through hinge-like motions.
Interestingly, it will also undergo a conformational change in the absence of substrates \cite{Henzler:2007} which makes the study of its transition particularly amenable to numerical simulation because it removes the complexity of accurately simulating the protein –ligand interactions.
As such, AKE's transition between its open and closed states  has been widely studied both by computational simulations (see for example Ref. \cite{Seyler:2014} and references therein), making it a perfect illustrative example for any method generating transition paths.

 \begin{table*}[!hbt]
	\centering
        \begin{threeparttable}
        \caption{Predicting intermediates by constructing transition paths}
        \label{table:improvement}
                \begin{tabular}{l c c c c c c c c c c c c c c c c c c c c c c c}
                \cline{1-24} 
                & \multicolumn{3}{c}{PDB ID} & & \multicolumn{3}{c}{cRMS (\AA)} & & & \multicolumn{2}{c}{SIDE} &&& \multicolumn{2}{c}{CLD} &&&  \multicolumn{2}{c}{EBDIMS} &&& \multicolumn{2}{c}{MAP}\\
                \cline{2-4} \cline{6-8} \cline{11-12} \cline{15-16} \cline{19-20} \cline{23-24}
                Name & Start (A) & Intermediate (I) & Final (C) & & (AI) & (IC) & (AC) & & & R$_{best}$\tnote{a} & IS \tnote{b)} & & & R$_{best}$ & IS & & & R$_{best}$ & IS & & & R$_{best}$ & IS \\
                 \hline
                 5'-NT & 1oidA & 1oi8B & 1hpuD && 5.4 & 4.7 & 9.3 & & & 2.01 & 58.0 &&& 1.61 & 66.0 &&& 2.47 & 48.0 &&& 2.14 & 55.0 \\
                 RBP & 1ba2A & 1urpD & 2driA && 2.2 & 4.20 & 6.2 && & 0.82 & 63 &&& 0.77 & 65 &&& 1.76 & 20.0 &&& 0.70 & 68 \\
                 RNase III & 1yyoAB & 1yz9AB & 1yywAB && 7.3 & 13.2 & 17.5 && & 5.0 & 30.6 &&& NA \tnote{c} & NA &&& 5.4 & 25.5 &&& 6.8 & 6.2 \\
                 CA$^{2+}$-ATPase & 1su4A & 1vfpA & 1iwoA && 13.7 & 10.1 & 114.0 && & 9.3 & 7.4 &&& NA & NA &&& 8.9 & 11.9 &&& 9.4 & 7.6 \\
                 Myosin & 1qviA & 1kk7A & 1kk8A && 16.7 & 12.0 & 27.3 && & 4.2 & 65.0 &&& NA & NA &&& 5.9 & 50.6 &&& 4.2 & 65.0\\
                \hline
                 \end{tabular}
                \begin{tablenotes}
                \item [a)] {\small Minimal cRMS (over CA atoms, in Angstroms) between the trajectory and the intermediate state, computed using Equation \ref{eqn:is1}. }
		\item [b)] {\small Improvement score, in percent, computed using Equation \ref{eqn:is}. The higher the number, the better. The best score for each protein is highlighted in bold.}
		\item [c)] {\small Trajectory not found (see text)}
	         \end{tablenotes}
	         \end{threeparttable}
\end{table*}

 Figure \ref{fig:ake_traj} illustrates a trajectory generated by SIDE between the closed state (PDB entry 1Ake) and the open state (PDB entry 4ake) of AKE.  As expected, the main motion is a hinge motion that leads the LID to separate from the CORE and the NMP. 
 During the transition, the elastic network associated with the protein decreases in size. Initially, the protein is very compact and edges are observed within, and in between the three domains of the protein. As we get closer to the open conformation, the LID maintains only limited interactions with the two other domains; this is captured by the elastic network.
 
   We compare in Figure \ref{fig:ake_comp} the four trajectories generated by SIDE, CLD, MAP, and EBDIMS, using both geometry (panel A), and a projection onto the principal components of the space of conformations for AKE (panel B).
 Figure \ref{fig:ake_comp}A shows a significant transition in the angle  LID-CORE, $\theta_1$.
 It changes from 95 $^{\circ}$ to 60 $^{\circ}$, corresponding to the open and the closed state of LID domain, respectively.
In contrast, the angle of NMP-CORE, $\theta_2$, varies from 65 $^{circ}$ to 35 $^{circ}$, corresponding to the open and semi-open state of
the NMP domain, respectively.
Interestingly, the two angular transitions appear to occur simultaneously in all 4 trajectories. 

Figure \ref{fig:ake_comp}B is generated by projecting the four trajectories on the first two principal components (PC) of the structural ensemble generated by combining all snapshots along those trajectories. 
The two trajectories SIDE and CLD are very close to each other. Those two trajectories are based on the same equation of motions, but with two very distinct potentials. Interestingly, those two trajectories are ``ballistic" in the sense that they are relatively straight in structure space.

\begin{figure}[htb]
\centering
\includegraphics[width=0.3\textwidth]{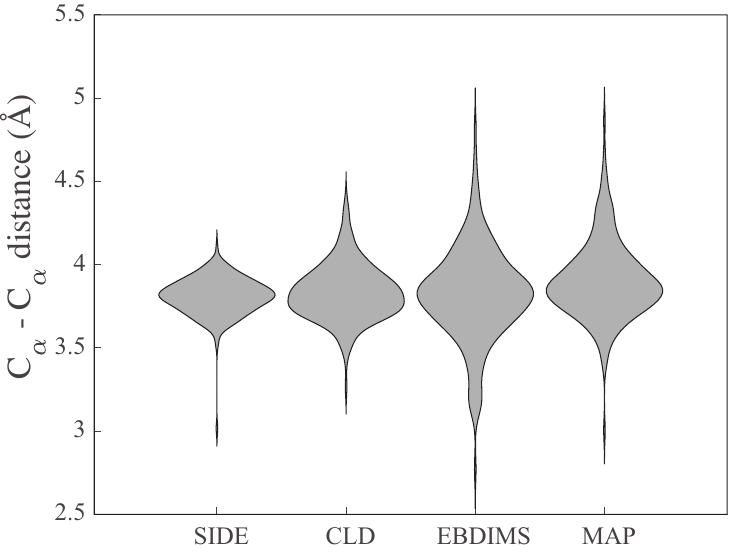}
\caption{The distributions of the distances between neighboring C$_\alpha$s for one snapshot along the trajectory between the open state and closed state of AKE are shown as violin plot. For each method, we picked the snapshot whose corresponding distribution has the largest variance. Note the presence of a peak in those distribution at the C$_\alpha$-C$_\alpha$ distance of $2.9 \AA$ corresponding to the cis Proline 87.}
\label{fig:ake_ca}
\end{figure}

The four methods we have tested rely on a coarse-grained representation of the protein structure that only considers the position of the C$_\alpha$ atoms of the molecule. As such, it is legitimate to ask how well the geometry of the molecule is maintained. 
For each method considered here, we computed the distributions of the distances between neighboring C$_\alpha$s for all snapshots
of the trajectory it generated.
In figure \ref{fig:ake_ca}, we plot the distributions at the snapshot whose corresponding variance is the largest, for all four methods considered.
Of the four trajectories, SIDE exhibits the smallest variance in those distributions. This is by far not unexpected, as the potential used by SIDE explicitly constrains the distances between neighboring C$_\alpha$s. 
The three other trajectories keep those distances close to the expected value of $3.8$ \AA, with the largest variance for the EBDIMS trajectory.

\begin{figure}[htb]
\centering
\includegraphics[width=0.3\textwidth]{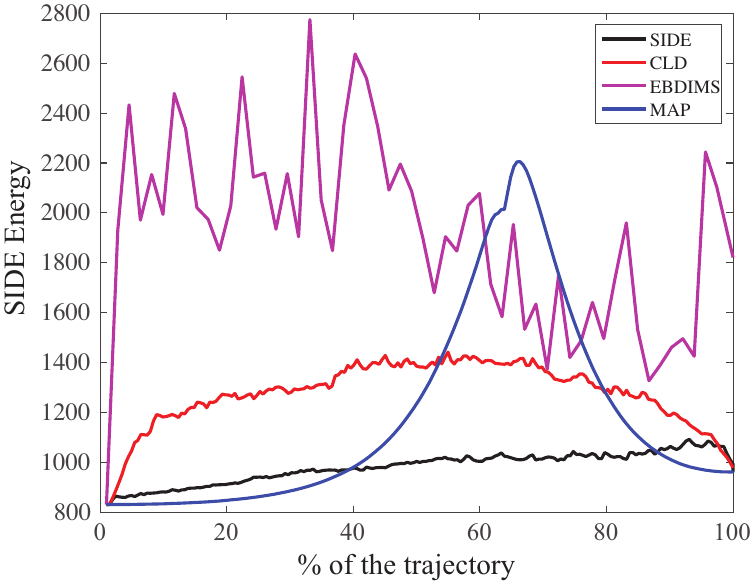}
\caption{\textbf{ The G\={o}-like energy along the  four  trajectories between the open state and closed state of AKE}, SIDE (black), CLD (red), EBDIMS (magenta) and MAP (blue).}
\label{fig:ake_ene}
\end{figure}   

In figure \ref{fig:ake_ene}, we plot the energy used by the SIDE strategy for all four trajectories between the open and closed states of AKE. 
This energy has three main components: a G\={o} like potential that maintains the geometry of the backbone of the protein by restraining the distances and angles between C$_{\alpha}$ atoms, a vdW term to prevent collision, and an elastic potential to drive the transition.
SIDE exhibits the lowest energy along the whole trajectory: this is not surprising, as it is explicitly the energy that drives its equation of motion.
Interestingly, however, this energy is relatively constant, hinting that the method was able to find a path that remains in low energy regions of the conformation space.
In contrast, the energy of the MAP trajectory shows a significant increase at a transition point. This is inherent to the method behind MAP that computes two quadratic trajectories from the two end points of the path and finds a transition point between those trajectories.
The energies along the EBDIMS trajectory are the largest, likely a consequence of the fact that the protein chain along this trajectory exhibits the largest deviation for standard geometry (see figure \ref{fig:ake_ca}).

%%%%%%%%%%%%%%%%%%%%%%%%%%%%%%%%%%%%%%%%%%%%%%%%%%%%%%%%%%%%%%%%%%%%%%%%%%%%%
\subsection{Detecting intermediate structures along transition paths}
%%%%%%%%%%%%%%%%%%%%%%%%%%%%%%%%%%%%%%%%%%%%%%%%%%%%%%%%%%%%%%%%%%%%%%%%%%%%%

Many methods have been developed to generate a path in conformational space between two structural states of a molecular system. As highlighted by Weiss and Levitt \cite{Weiss:2009}, however, there is no fully satisfying objective methods to test the biological relevance of such paths. Here we
follow the approximate method that they had proposed. Namely, we start from a set of proteins for which there are at least three distinct  conformations whose (experimental) structures are known and available in the Protein Data Bank \cite{Berman:2000}. Two of those conformations, $A$ and $C$ serve as end points for the transition, while the third is set to be the intermediate state, $I$; the distinction between the three comes from biology. Trajectories between the two end points are generated, with no knowledge of the intermediate state. We then follow how close the trajectory comes to the intermediate structure, 
\begin{eqnarray}
R_{best}(I) = \min_{k\in[1,N]}[cRMS(C_k,I)],
\label{eqn:is1}
\end{eqnarray}
where the min  is taken over all conformations $C_k$ along the trajectory, and by computing an improvement score IS \cite{Weiss:2009, Tekpinar:2010}:
\begin{eqnarray}
IS = 100 \times \left( 1 - \frac{\displaystyle R_{best}(I)}{\min(cRMS(A,I),cRMS(C,I))} \right),
\label{eqn:is}
\end{eqnarray}
As defined, $IS$ is a measure of how close the trajectory comes to the structure of the intermediate, computed as a fraction of how close the start and end points are to this intermediate. In both equations \ref{eqn:is1} and \ref{eqn:is}, cRMS stands for the coordinate root mean square deviation computed over all C$_{\alpha}$s of the protein.

\begin{figure*}[htb]
\centering
\includegraphics[width=0.6\textwidth]{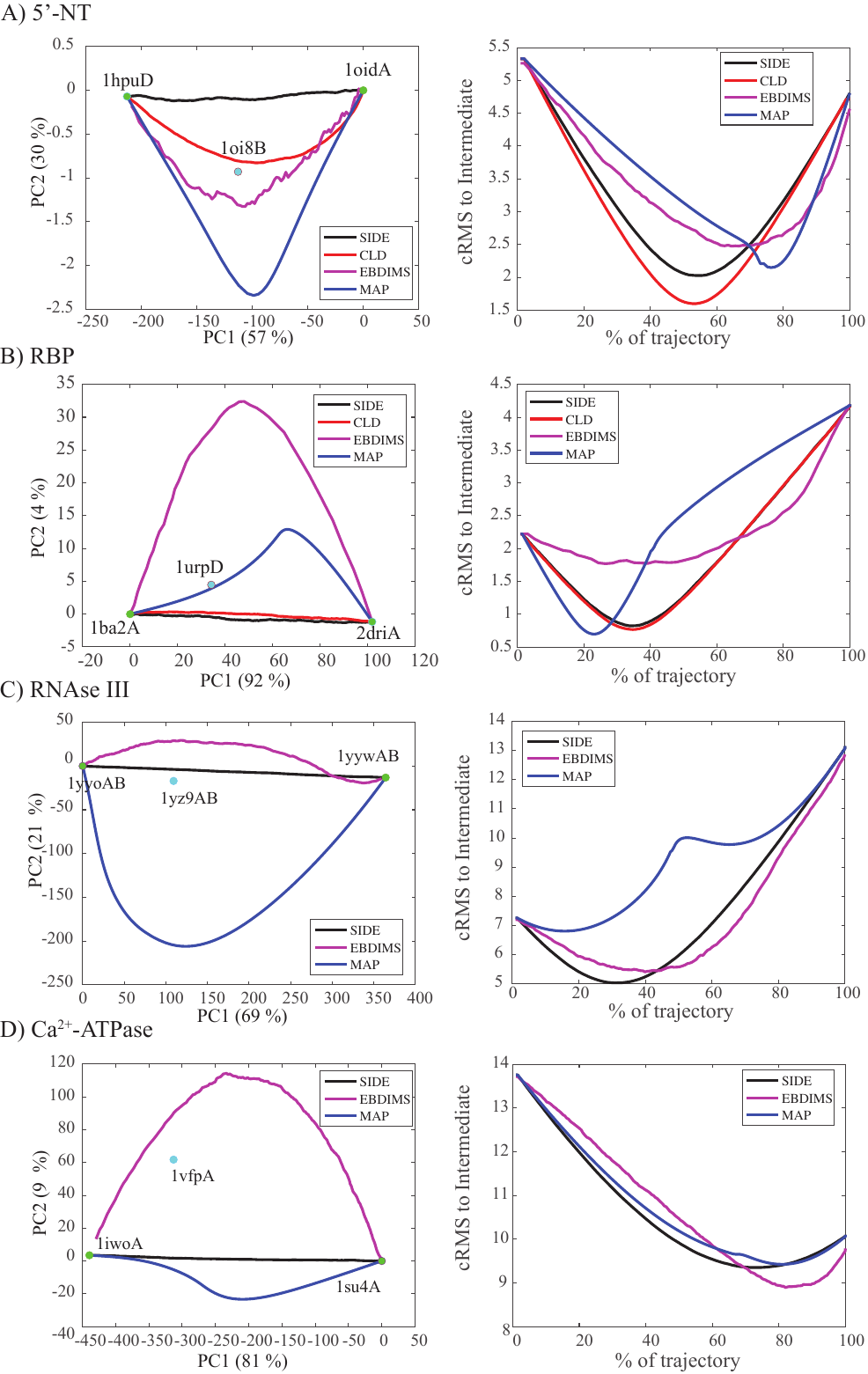}
\caption{We computed trajectories between two states A and C for four proteins (5'-NT, RBP,  RNAse, and Ca$^{2+}$-ATPase) using SIDE (black), CLD (red), EBDIMS (magenta) and MAP (blue). Right panels, projections of the 4 trajectories along the 2 principal components of their conformational spaces, and, left panels, distance of a putative intermediate I to successive snapshots of the trajectory. CLD trajectories for RNAse III and Ca$^{2+}$-ATPase and missing as we could not find parameters to get them to converge.}
\label{fig:intermediates1}
\end{figure*}

\begin{figure*}[htb]
\centering
\includegraphics[width=0.6\textwidth]{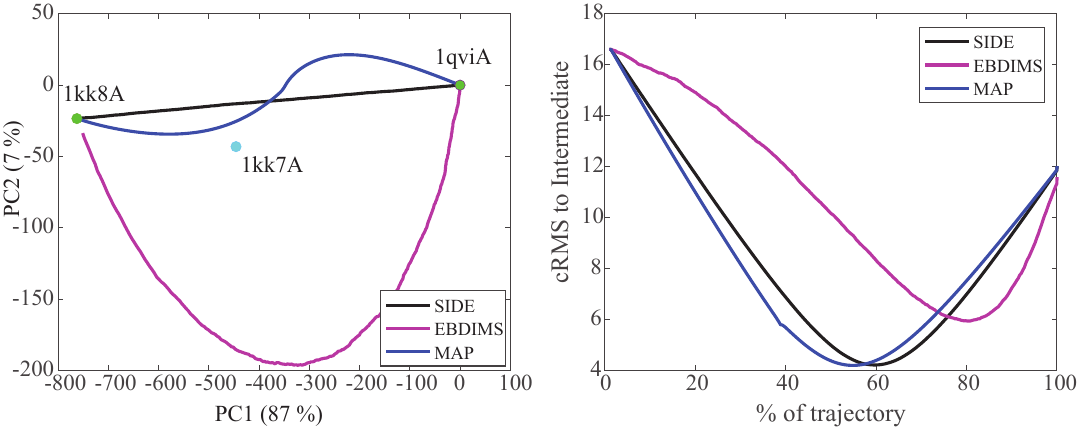}
\caption{We computed trajectories between two states of  scallop myosin II (1qviA and 1kk8A)  using SIDE (black), EBDIMS (magenta) and MAP (blue). Right panel, projections of the 4 trajectories along the 2 principal components of their conformational spaces, and, left panels, distance of a putative intermediate 1kk7A to successive snapshots of the trajectory.}
\label{fig:intermediate2}
\end{figure*}

Results for five proteins included in the original paper from Weiss and Levitt and the four methods for generating transition paths considered in this study are shown in Figures \ref{fig:intermediates1}, \ref{fig:intermediate2}, and Table \ref{table:improvement}.

\subsubsection{Four ``simple" transitions}

The RBP protein corresponds to the easiest of the four cases considered here: the motion between the start and final states is a simple hinging motion. SIDE, CLD, and MAP perform very similarly in terms of IS score.  All three methods get within one Angstrom of the target intermediate structure (1urpD) . Interestingly, in the PC space, the intermediate structure is closer to the MAP trajectory, while the IS score would indicate that the CLD trajectory gets closer. EBDIMS does not seem to identify a trajectory that gets close to the intermediate. This could be just a question of tuning its parameters.

The 5'-NT protein undergoes a large domain rotation. All four methods are able to capture that motion, with the resulting trajectories getting close to the intermediate conformation. Based on IS, CLD performs better (as indicated by the Improvement Score and minimal cRMS), with its trajectory getting closer than 1.65 \AA\ from the intermediate state, for an improvement score of 66\%. In contrast to the RBP test case, this observation is confirmed in the PC space.

The RNAse III protein undergoes much larger conformational changes than both RBP and 5'-NT as the protein switches from the catalytic form to the non-catalytic form.
In this switch, the orientations of the two domains of each of the two chains of RNAse are changed drastically. 
CLD fails to generate such a trajectory; this was already noted in our previous study \cite{Delarue:2017}. Both SIDE and EBDIMS perform well on this protein, generating smooth transitions that gets within 5 \AA\ of the intermediate conformation, with improvement scores of 30 \%. and 26\%, respectively.  In comparison, MAP fails to get near the intermediate conformation (improvement score of 6.2 \%). Note that in PC space, the SIDE and EBDIMS trajectories are very close to each other.

The Ca$^{2+}$ ATPase is the most complicated of the four cases illustrated in Figure \ref{fig:intermediates1} . The transition between its start and final conformations (apo and holo conformations, respectively) involve a significant structural rearrangement (14 \AA). CLD again failed to generate a trajectory. None of the other three methods  capture correctly this transition, as none get significantly closer to the intermediate conformation (they do get closer, but stay below 9 \AA of the intermediate). This failure should be considered as relative, as the transition involve a complicated series of conformational changes  \cite{Toyoshima:2002}. We note that Mixed-ENM seems to perform best, with an improvement score of 8.5 \%, but its trajectory remains more then 9 \AA\ away from the intermediate conformation.  

\subsubsection{A more difficult test case: Myosin}

Weiss and Levitt considered a fifth test case, scallop myosin, that is often not included in subsequent studies, as it is the most challenging of the test cases they considered, with the largest structural distance between the two conformations considered (1qviA and 1kk8A), and between those two end points and the structure of the putative intermediate, 1kk7A (see table \ref{table:improvement}). We analyze it in more detail here.

Myosin is a motor protein, hydrolyzing ATP to drive muscle contraction. Weiss and Levitt built a trajectory between 
pre-stroke and post-stroke structures, using the near-rigor structure as the intermediate; this corresponds to the order in which they are found in the power cycle \cite{Himmel:2002}.

We built three trajectories between the pre-stroke (1qviA) and post-stroke (1kk8A) of scallop myosin, using SIDE, EBDIMS, and MAP. Just like for RNAse III and Ca$^{2+}$ ATPase, we could not find parameters that would allow us to generate a trajectory with CLD. In Figure \ref{fig:intermediate2}, we illustrate both the projections of the three trajectories in the PC space of the structural ensemble obtained by combining them, as well as the distances of the intermediate 1kk7A to successive snapshots of the trajectories. 
SIDE and MAP  perform well on this protein, generating smooth transitions that get within 4.2 \AA\ of the intermediate conformation, with improvement scores of 65 \% (see table \ref{table:improvement}). 
The trajectory generated by EBDIMS is significantly different (left panel of Figure \ref{fig:intermediate2}); it also
get closer to the intermediate conformation, albeit to a lesser extent (IS of 50\%).

\begin{figure*}[htb]
\centering
\includegraphics[width=0.6\textwidth]{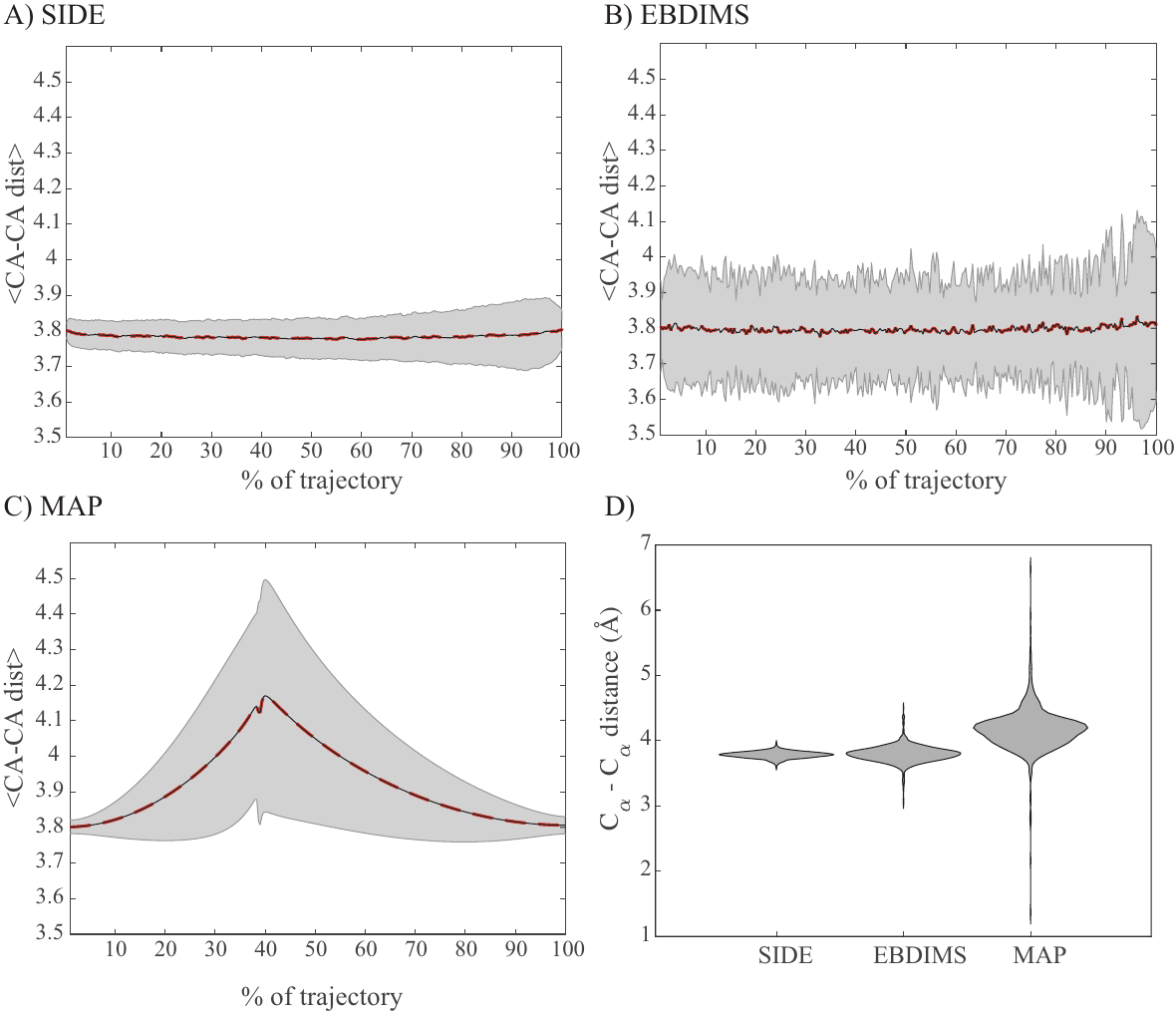}
\caption{The mean (dotted line) and standard deviation (grey shaded region) of the distributions of the distances between neighboring C$_\alpha$s for all snapshots along the SIDE (panel A), EBDIMS (panel B), and MAP (panel C) between the pre-stroke and post-stroke structures of scallop myosin. In panel  (D) the full distribution is shown as a violin plot for one snapshot along each trajectory. For each method, we picked the snapshot whose corresponding distribution has the largest variance.}
\label{fig:myosin_ca}
\end{figure*}   

The results illustrated in Figure \ref{fig:intermediate2}, however, could be misleading as each data point on the figures summarizes a whole protein structure. Both 1qviA and 1kk8A have a few missing residues within loops that may impact the quality of the trajectories designed by the three methods considered. In Figure \ref{fig:myosin_ca}, we show the distributions of the distances between neighboring C$_\alpha$s for all snapshots of the trajectories (panels A, B, and C), as well as specific distributions at the snapshots corresponding to the distributions with the largest variances (panel D). The SIDE trajectory is very well constrained, maintaining CA-CA distances within 0.1 \AA\ or the ideal value of 3.8 \AA. The EBDIMS trajectory shows more variations, with some CA-CA distances close to 5 \AA. The most striking behavior, however, is observed for the MAP trajectory, exhibiting large variations near its transition point. Some CA-CA distances are found close to 1 \AA, while others are close to 7 \AA.
As such, even though the trajectory seems reasonable from Figure \ref{fig:intermediate2}, it should be difficult to use it to generate full atom models for its snapshots.

\subsubsection{Overall assessment}

The trajectories generated by SIDE, CLD, EBDIMS, and MAP all have merits in their abilities to identify motions responsible to changes of conformation as well as in their capacity to retrieve intermediate structures. There are,
however, differences that are worth summarizing.
CLD and MAP work well for systems with small cRMS between the two conformations at the end points of
the trajectories. 
When the cRMS becomes large (for example, for RNAse III or myosin), CLD fails (more exactly, we could not find parameters to make it work), while MAP ``explodes", i.e., generates trajectories with severe geometric problems. This was already observed before \cite{Koehl:2016}.
EBDIMS is more consistent, generating trajectories in all 5 test cases we considered. It does show some distortions along the main chain of the protein.
In contrast, SIDE performs well on all 5 cases, leading to trajectories whose snapshots have correct stereochemistry.
The difference between EBDIMS and SIDE is likely due to the difference in their potential, as the latter defines the protein main chain geometry explicitly.

%%%%%%%%%%%%%%%%%%%%%%%%%%%%%%%%%%%%%%%%%%%%%%%%%%%%%%%%%%%%%%%%%%%%%%%%%%%%%
\subsection{Limitation of path sampling methods: the VATP case}
%%%%%%%%%%%%%%%%%%%%%%%%%%%%%%%%%%%%%%%%%%%%%%%%%%%%%%%%%%%%%%%%%%%%%%%%%%%%%

The examples presented above are standard test cases that have been used in many studies. 
While useful in highlighting strengths and weaknesses of path sampling methods, they fail to report on a critical issue for coarse-grained approaches, that is illustrated in the following.

ATP synthase enzymes are ``splendid molecular machines" \cite{Boyer:1997,  Junge:2015} that catalyze the synthesis of ATP. These enzymes exist in two main families: F-type ATP synthases, found in mitochondria, chloroplasts, and bacterial membranes, and V-type ATPases, predominantly located in vacuolar membranes and specialized cellular compartments. Both types share a common architectural principle featuring two coupled rotary motors. The catalytic mechanism fundamentally depends on a rotary motion where proton (or ion) flow through a membrane-embedded rotor domain drives the physical rotation of a central stalk. This mechanical rotation is then transmitted to the catalytic domain, inducing conformational changes in the active sites that enable ATP synthesis in F-type enzymes or ATP hydrolysis to drive proton pumping in V-type enzymes. 
Understanding this rotary motion and its coupling to the catalysis/hydrolysis reactions is therefore essential to the comprehension of how this enzyme works \cite{Nakamoto:2008}.

\begin{figure}[htb]
\centering
\includegraphics[width=0.2\textwidth]{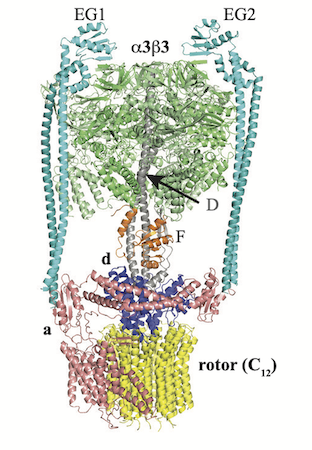}
\caption{\textbf{The V-ATPase of \emph{Thermus Thermophilus} in the low energy rotational state 1}, PDB code 6qum, color-coded as subunits $\alpha$ and $\beta$ in green, subunit D is grey, F in orange, the two peripheral staks EG1 and EG2 is teal and cyan, respectively, domain d in blue, a in salmon, and the rotor formed by 12 helices in yellow.}
\label{fig:synthase}
\end{figure}   

\begin{figure*}[htb]
\centering
\includegraphics[width=0.80\textwidth]{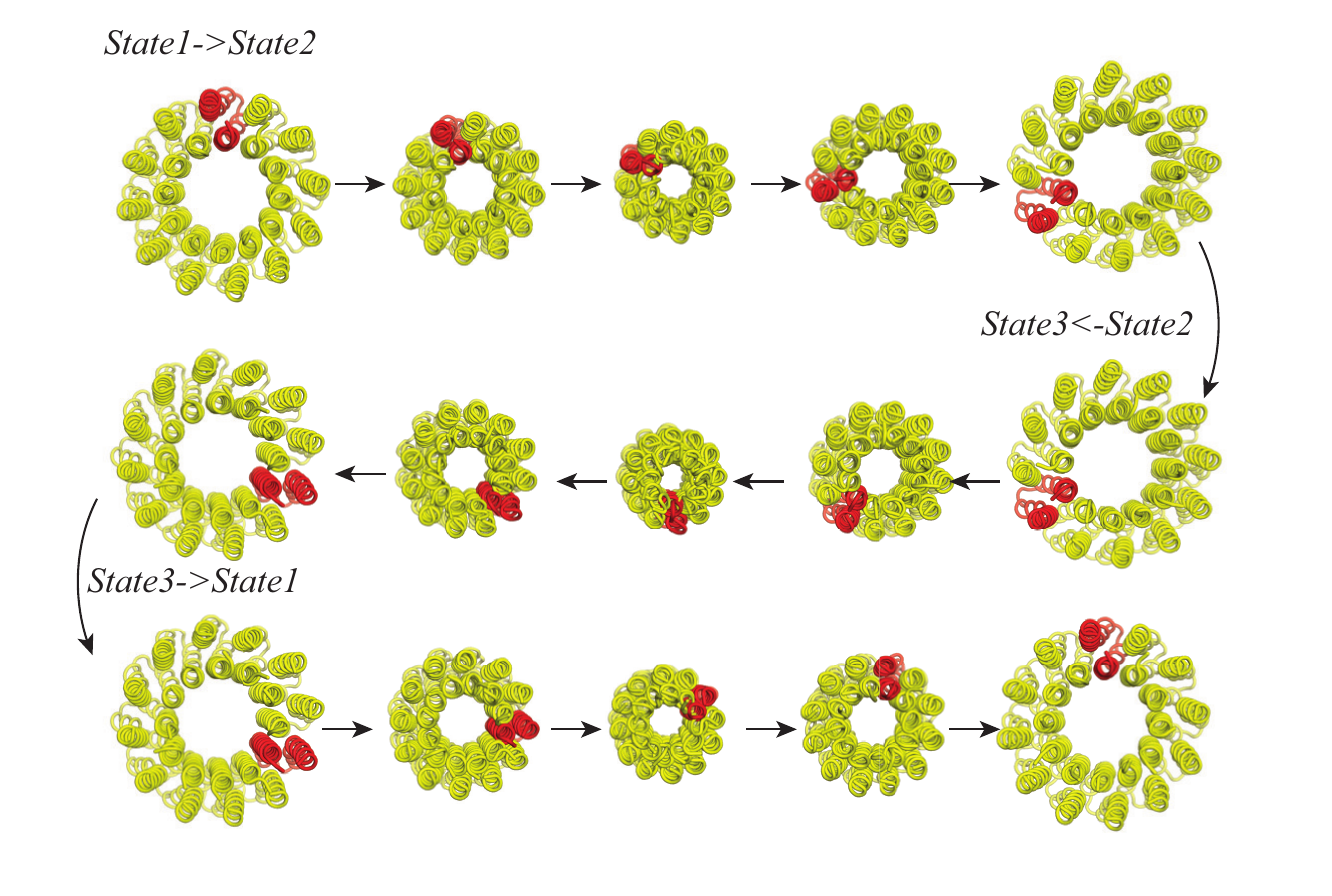}
\caption{\textbf{Rotation of the rotor of The V-ATPase of \emph{Thermus Thermophilus}}. We plot bottom views (i.e. from the cytoplasm) of the rotor of VATP along the full trajectory reconstituted from the three trajectories state 1 (PDB 6qum) -> state 2 (PDB 6r0w), state 2->state 3 (PDB 6r0y), and state 3->state 1. One of the 12 helices of the rotor is colored in red to illustrate the rotational motion. All images are at the same scale: note the unusual breathing motion of the whole rotor between each end state of the trajectories. }
\label{fig:rotor}
\end{figure*}   

Here we consider the V-type ATPase (VATP) from \emph{Thermus Thermophilus}. Cryo-EM data revealed three main conformations for the enzyme, corresponding to three states along the full rotation of the rotor at 120$^{\circ}$ apart \cite{Zhou:2019}. Figure \ref{fig:synthase} illustrates the architecture of one of those states. The three conformations for VATP are available in the PDB with ids 6qum (state 1), 6r0w (state 2), and 6r0y (state 3). We built three trajectories, S12 (state 1 $\rightarrow$ state 2), S23 (state 2  $\rightarrow$ state 3), and S31 (state 3 $\rightarrow$ state 1), using SIDE, that would ultimately reconstitute the whole rotation of the rotor. Results illustrating this complete rotation at the level of the rotor are shown in Figure \ref{fig:rotor}.

\begin{figure}[htb]
\centering
\includegraphics[width=0.30\textwidth]{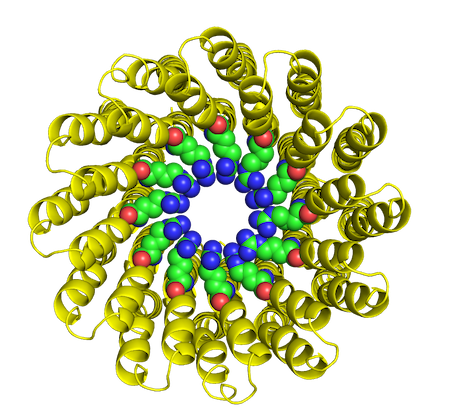}
\caption{\textbf{The rotor of V-ATPase of \emph{Thermus Thermophilus} in the low energy rotational state 1}, (PDB code 6qum). The 12 helices forming the barrel of the rotor are colored yellow. We highlight the corresponding 12 arginine at position 36 that extends within the lumen of the rotor.}
\label{fig:vatp_arg}
\end{figure}

Each of the 3 trajectories, S12, S13, S31, captures the corresponding rotation of the rotor. The geometry of the individual helices is well preserved. However, we observe a breathing motion, leading to a reduction in size of the whole rotor, along each of the trajectories. This breathing motion is not expected to be natural. For example, it is unlikely due to the presence of side chains in the lumen of the rotor, as illustrated in Figure \ref{fig:vatp_arg}. The problem is that the potential used by SIDE knows nothing about those side chains. It is based on a coarse-grained representation of the protein in which only the C$_{\alpha}$ atoms are considered. It includes terms to maintain the geometry of the backbone. Its collision term can't avoid the breathing motion we observe, as atoms within the side chains are not there. This is a general problem. We ran similar simulations with CLD (failed) and with MAP and EBDIMS. The corresponding trajectories exhibit similar non-natural motions, even distortions of the helices of the rotor.

%%%%%%%%%%%%%%%%%%%%%%%%%%%%%%%%%%%%%%%%%%%%%%%%%%%%%%%%%%%%%%%%%%%%%%%%%%%%
%%%%%%%%%%%%%%%%%%%%%%%%%%%%%%%%%%%%%%%%%%%%%%%%%%%%%%%%%%%%%%%%%%%%%%%%%%%%
\section{Conclusions and Perspective}
%%%%%%%%%%%%%%%%%%%%%%%%%%%%%%%%%%%%%%%%%%%%%%%%%%%%%%%%%%%%%%%%%%%%%%%%%%%%
%%%%%%%%%%%%%%%%%%%%%%%%%%%%%%%%%%%%%%%%%%%%%%%%%%%%%%%%%%%%%%%%%%%%%%%%%%%%

In this paper, we addressed the problem of generating paths for a bio-molecular system that start at a given initial configuration and that are conditioned to end at a given final configuration.
Our approach follows the ideas of Langevin overdamped dynamics, as expressed with  the bridge equation  \cite{Orland:2011, Majumdar:2015, Delarue:2017}.  
We first revisited this concept of bridge in the context of low temperature, leading to conditioned Langevin dynamics, already described in earlier work \cite{Delarue:2017, Koehl:2022}.
We introduced a new coarse-grained potential that describes the stability of the system and illustrated 
how the combination of conditioned Langevin dynamics with this new potential in a framework we call SIDE
can generate realistic transition pathways for proteins.
We tested SIDE against previous iterations of our efforts to generate transition paths as well as against publicly available tools to compute such tasks.
On a test set of 5 proteins originally introduced by Weiss and Levitt \cite{Weiss:2009}, we showed that SIDE  performs as well as all the other methods tested, with improved conservation of the protein backbone geometry along the trajectory.
Finally, we highlight a limitation that is not inherent to our method, but associated with the use of coarse-grained representations of proteins.

Methods that generate trajectories between two conformations of the same protein are defined with two main components: their equation of motion,
and the potential they use to represent the energetics of the protein.
We tried to disentangle the two by comparing SIDE, our new framework, with CLD, our previous \cite{Delarue:2017} path sampling technique, as those two methods
rely on exactly the same equation of motion, a conditioned Langevin bridge, with the same approximation of low temperature.
CLD relies on a mixed-ENM potential that combines information from the two conformations (start and end), while SIDE relies on a pseudo-elastic potential with no (or zero) reference state. CLD adds a repulsive collision term, while SIDE adds both a vdW collision term and a geometric potential to maintain correct main chain geometry along the trajectory. We found that applications of CLD are limited to systems whose transitions involve relatively small changes or simple motion, such as a hinge motion,
as it is difficult to parameterize the mixed-ENM potential (this was already observed before \cite{Delarue:2017}). 
SIDE is much easier to parameterize, at least when it comes to defining the potential, and was successful in (nearly) all the test cases considered (where ``nearly" is explained below).

We have provided evidence that SIDE is able to predict intermediate structures in the transition between two conformations of a protein, based on the knowledge of those two conformations only, in addition to generating 
geometrically realistic trajectories. 
While promising, those successes should still be considered with caution. The most successful applications described here correspond to cases in which the motions involved in the transition are relatively simple, such as hinge motions or domain rotations. For more complicated motions, such as those involved in the test case 
$Ca^{2+}$ ATPase or VATP,  SIDE (and all other methods tested), perform poorly. 
Of greater concern, SIDE can generate trajectories with unrealistic motions, such as the breathing found for the rotor of VATP.
The latter is clearly associated with limitations of a coarse-grained potential that only considers one atom per residue. While such potentials are useful to reduce computational costs, enabling simulations of large systems,  the
limitation we illustrate is of concern.
One solution to address this problem is to add constraints to the potential. For example, in the case of VATP, we could have introduced constraints to keep the helices of the rotors distributed around a barrel of (nearly) fixed size. This type of ad-hoc solution, however, is difficult to implement as it is not always clear what those
constraints should be.
Another approach is to refine the description of the protein. 
We are currently investigating the use of more realistic coarse-grained representations of protein than the C$_{\alpha}$ only representation, such as the MARTINI force field \cite{Marrink:2007, Monticelli:2008}, more specifically the promising G\={o}Martini3 model \cite{Souza:2025} that combines G\={o} models \cite{Taketomi:1975} with the Martini3 force field \cite{Souza:2021}.

Finally, it is worth mentioning that while all the results presented here relate to protein structural transitions, there is nothing in the equation of motion or even the potential that would prevent applications to nucleic acids, both RNAs and DNAs, as well as to protein-nucleic acids complexes.

%%%%%%%%%%%%%%%%%%%%%%%%%%%%%%%%%%%%%%%%%%%%%%%%%%%%%%%%%%%%
%%%%%%%%%%%%%%%%%%%%%%%%%%%%%%%%%%%%%%%%%%%%%%%%%%%%%%%%%%%%
\begin{acknowledgments}
The work discussed here originated from a visit by P.K. at the Institut de Physique Th\'{e}orique, CEA Saclay, France, during the fall of 2025. He thanks them for their hospitality and financial support.
\end{acknowledgments}
%%%%%%%%%%%%%%%%%%%%%%%%%%%%%%%%%%%%%%%%%%%%%%%%%%%%%%%%%%%%
%%%%%%%%%%%%%%%%%%%%%%%%%%%%%%%%%%%%%%%%%%%%%%%%%%%%%%%%%%%%

%\section*{Data Availability Statement}

%%%%%%%%%%%%%%%%%%%%%%%%%%%%%%%%%%%%%%%%%%%%%%%%%%%%%%%%%%%%
%%%%%%%%%%%%%%%%%%%%%%%%%%%%%%%%%%%%%%%%%%%%%%%%%%%%%%%%%%%%

%The data that support the findings of this study are openly available in the PDB repository at the URL \url{https://www.rcsb.org/}

\appendix

%%%%%%%%%%%%%%%%%%%%%%%%%%%%%%%%%%%%%%%%%%%%%%%%%%%%%%%%%%%%
%%%%%%%%%%%%%%%%%%%%%%%%%%%%%%%%%%%%%%%%%%%%%%%%%%%%%%%%%%%%
\section{Pairwise potential: Gradient, Hessian, and Laplacian}
\label{sec:A1}
%%%%%%%%%%%%%%%%%%%%%%%%%%%%%%%%%%%%%%%%%%%%%%%%%%%%%%%%%%%%
%%%%%%%%%%%%%%%%%%%%%%%%%%%%%%%%%%%%%%%%%%%%%%%%%%%%%%%%%%%%

%%%%%%%%%%%%%%%%%%%%%%%%%%%%%%%%%%%%%%%%%%%%%%%%%%%%%%%%%%%%
\subsection{Notations}
%%%%%%%%%%%%%%%%%%%%%%%%%%%%%%%%%%%%%%%%%%%%%%%%%%%%%%%%%%%%
 
Let $P$ be a protein consisting of $N$ atoms, with atom $i$ characterized by its position $\mathbf{r}_i$.
The whole molecule is described by a $3N$ position vector $\mathbf{X} = (\mathbf{r}_1, \mathbf{r}_2, \ldots, \mathbf{r}_N)^T \in \mathbb{R}^{3N}$.
We set $\mathbf{X}^0$ to be the start conformation of the system.
For two atoms $i$ and $j$ of $P$, we set $r_{ij} =|\mathbf{r}_i-\mathbf{r}_j|$ and $r_{ij}^0=|\mathbf{r}_i^0-\mathbf{r}_j^0|$ to be the Euclidean distances between them
 in a conformation $\mathbf{X}$ and in the ground-state conformation $\mathbf{X}^0$, respectively.  Similarly, we define $\mathbf{r}_{ij}=\mathbf{r}_i-\mathbf{r}_j$ and $\mathbf{r}_{ij}^0=\mathbf{r}_i^0-\mathbf{r}_j^0$ the vectors along the edge (i,j) in $\mathbf{X}$ and in $\mathbf{X}^0$, respectively.
The unit vector along the same edge is referred to as $\mathbf{e}_{ij}$ and is equal to 
\begin{eqnarray}
e_{ij} = \frac{\mathbf{r}_{ij}}{r_{ij}}
\end{eqnarray}
We write $I_3$ for the $3\times3$ identity matrix.

%%%%%%%%%%%%%%%%%%%%%%%%%%%%%%%%%%%%%%%%%%%%%%%%%%%%%%%%%%%%
\subsection{A generic pairwise potential}
%%%%%%%%%%%%%%%%%%%%%%%%%%%%%%%%%%%%%%%%%%%%%%%%%%%%%%%%%%%%

For a pair atoms $(i,j)$ we define
\begin{eqnarray}
U_{ij}  = f(r_{ij})
\label{eqn:pair}
\end{eqnarray}
$f$ is a function that is $C^\infty$ almost everywhere.
This general formulation applies to the bond term, the collision term, and the elastic term of the full potential we have used to study conformational transitions in proteins. Note that in the case of the collision potential, $f$ is not defined for $r_{ij}=0$.

%%%%%%%%%%%%%%%%%%%%%%%%%%%%%%%%%%%%%%%%%%%%%%%%%%%%%%%%%%%%
\subsection{Gradient}
%%%%%%%%%%%%%%%%%%%%%%%%%%%%%%%%%%%%%%%%%%%%%%%%%%%%%%%%%%%%

It is convenient to define the  vector $\mathbf{W}_{ij}$
\begin{eqnarray}
\mathbf{W}_{ij}=(0,\ldots,0,\mathbf{e}_{ij},0,\ldots,0,-\mathbf{e}_{ij},0,\ldots,0),
\end{eqnarray}
namely, $\mathbf{W}_{ij}$ is a vector in $\mathbb{R}^{3N}$ that is zero everywhere, except at positions $i$ and $j$ where it is equal to the unit vector $\mathbf{e}_{ij}$ and its opposite, respectively. We have,
\begin{eqnarray}
\nabla U_{ij} = f'(r_{ij}) \mathbf{W}_{ij}
\end{eqnarray}

%%%%%%%%%%%%%%%%%%%%%%%%%%%%%%%%%%%%%%%%%%%%%%%%%%%%%%%%%%%%
\subsection{Hessian}
%%%%%%%%%%%%%%%%%%%%%%%%%%%%%%%%%%%%%%%%%%%%%%%%%%%%%%%%%%%%

When we take second derivatives with respect to the Cartesian coordinates of particles \(i\) and \(j\),
the Hessian of the pair contribution \(U_{ij}\) is a \(3N\times3N\) matrix which we write in \(3\times3\)-block form as
\begin{eqnarray} 
H =
\begin{bmatrix}
    0      & \cdots & 0      & \cdots & 0     & \cdots & 0 \\
    \vdots & \ddots & \vdots &        & \vdots &         & \vdots \\
    0      & \cdots & H(i,i) & \cdots & H(i,j) & \cdots & 0\\
    \vdots &     & \vdots &  \ddots  & \vdots &         & \vdots \\
    0      & \cdots & H(j,i) & \cdots & H(j,j)  & \cdots & 0\\
     \vdots &     & \vdots &        & \vdots &  \ddots  & \vdots \\
     0      & \cdots & 0      & \cdots & 0     & \cdots & 0 \\
\end{bmatrix}
\end{eqnarray}

Expressions for the different blocks in this matrix are obtained by differentiating the gradient .
Notice first that
\begin{eqnarray}
\nabla_{\mathbf{r}_i} \mathbf{e}_i = \frac{1}{r_{ij}} \left( I_3 - \mathbf{e}_{ij} \mathbf{e}_{ij}^{T}\right).
\end{eqnarray}
Then,
\begin{eqnarray} 
H(i,i) &=& f''(r) \mathbf{e}_{ij} \mathbf{e}_{ij}^{T} + \frac{f'(r_{ij})}{r_{ij}} ( I_3 - \mathbf{e}_{ij} \mathbf{e}_{ij}^{T}).
\label{eq:Hii}
\end{eqnarray}
As $U_{ij}$ depends only on relative position, and not absolute position, it is easy to verify that
\begin{eqnarray} 
H(j,i) &=& H_(i,j) = -H(i,i)\nonumber \\
H(j,j) &=&H(i,i).
\end{eqnarray}

%%%%%%%%%%%%%%%%%%%%%%%%%%%%%%%%%%%%%%%%%%%%%%%%%%%%%%%%%%%%
\subsection{Laplacian and its gradient }
%%%%%%%%%%%%%%%%%%%%%%%%%%%%%%%%%%%%%%%%%%%%%%%%%%%%%%%%%%%%

As the trace is a linear operator, we get
\begin{eqnarray} 
\operatorname{tr}\big(H_{ii}\big)
&=&  f''(r_{ij})  \operatorname{tr}(\mathbf{e}_{ij}\mathbf{e}_{ij}^T)  + \frac{f'(r_{ij})}{r_{ij}}\operatorname{tr}(I_3 - \mathbf{e}_{ij} \mathbf{e}_{ij}^{T}) \nonumber \\
&=& f''(r_{ij}) + 2\frac{f'(r_{ij})}{r_{ij}}.
\end{eqnarray}

As the full trace involves both $H_{ii}$ and $H_{jj}$ and that these two blocks are equal, the trace of the Hessian H associated with the pair $(i,j)$ is:
\begin{eqnarray} 
\Delta U_{ij} = 2f''(r_{ij})+ 4\frac{f'(r_{ij})}{r_{ij}}.
\end{eqnarray}
Note that the gradient of the Laplacian is then:
\begin{eqnarray}
\nabla\left( \Delta U_{ij} \right) = \left( 2f'''(r_{ij})+ 4\frac{f''(r_{ij})}{r_{ij}} - 4\frac{f'(r_{ij})}{r^2_{ij}} \right) \mathbf{W}_{ij} \nonumber \\
\end{eqnarray}

%%%%%%%%%%%%%%%%%%%%%%%%%%%%%%%%%%%%%%%%%%%%%%%%%%%%%%%%%%%%
\subsection{Complete pairwise potentials: bond, collision, and elastic} 
%%%%%%%%%%%%%%%%%%%%%%%%%%%%%%%%%%%%%%%%%%%%%%%%%%%%%%%%%%%%

A potential associated with  the whole protein is the sum of the inter atomic interactions over a preset list of pairs $S$:
\begin{eqnarray}
 U=\sum_{(i,j)\in S} f(r_{ij}).
 \end{eqnarray}
 Linear superposition applies and we can just sum the gradients, Hessian, and Laplacian.
 
 Let us consider a single continuous chain with $N$ $C_{\alpha}$ atoms. The ``bond" potential is designed to constrain the links between consecutive $C_{\alpha}$s, i.e.,, to maintain their lengths as constant as possible. It is defined as:
 \begin{eqnarray}
 U_b &=& k_b \sum_{i=1}^{N-1} (r_{ij} - r_{ij}^0)^2,
 \end{eqnarray}
where we have set $j=i+1$ and $k_b = 100 \epsilon_G$. 
This pairwise potential maps exactly to the generic case considered above, using $f(r_ij)=(r_{ij}-r_{ij}^0)^2$ with $r_{ij}^0$ being a constant. 
Its gradient, Hessian, and Laplacian are then derived from the equations above:
 \begin{eqnarray}
 \nabla U_b &=& 2 k_b \sum_{i}  (r_{ij} - r_{ij}^0) W_{ij} \nonumber \\
 H_b(i,j) &=& -2 k_b \mathbf{e}_{ij} \mathbf{e}_{ij}^{T}  \nonumber \\
 && - \frac{2k_b(r_{ij} - r_{ij}^0)}{r_{ij}} (I_3 - \mathbf{e}_{ij} \mathbf{e}_{ij}^{T}) \nonumber \\
 H_b(i,i) &=& - \sum_{j \in \{i-1, i+1\}}  H_b(i,j) \nonumber \\
  \Delta(U_b) &=&   4 k_b \sum_{i} \sum_{j \in \{i-1, i+1\}} (3-2 \frac{r_{ij}^0}{r_{ij}} ) \nonumber \\
  \nabla\left( \Delta U_{b} \right) &=&   -8 k_b \sum_{i} \sum_{j \in \{i-1, i+1\}}  \frac{r_{ij}^0}{r_{ij}} W_{ij}
 \end{eqnarray}
Note that special care is needed for the first and last atoms of the chain.

The collision potential is a 12-6 Lennard Jones potential set to reduce the number of collision during the dynamics., It is defined as 
 \begin{eqnarray}
 U_{vdW} &=& \sum_{i} \sum_{j \in N(i)} \left( \frac{C_1}{r_{ij}^{12}}  - \frac{C_2}{r_{ij}^{6}} \right),
 \end{eqnarray}
 where $N(i)$ defines the neighborhood of $i$, i.e., the list of atoms that are within a cutoff distance $R_c$ of $i$), and $C_1=C_2=\epsilon_G$. 
 This pairwise potential maps also to the generic case considered above, using $f(r_ij)=\frac{C_1}{r_{ij}^{12}}  - \frac{C_2}{r_{ij}^{6}}$. 
 Its gradient, Hessian, and Laplacian are derived from the equations above:
 \begin{eqnarray}
 \nabla U_{vdW} &=& \sum_{i} \sum_{j \in N(i)} \left( \frac{-12C_1}{r_{ij}^{13}}  + \frac{6C_2}{r_{ij}^{7}} \right) W_{ij} \nonumber \\
 H_{vdW}(i,j) &=& -\left(\frac{156C_1}{r_{ij}^{14}}  - \frac{42C_2}{r_{ij}^{8}} \right) \mathbf{e}_{ij} \mathbf{e}_{ij}^{T}  \nonumber \\
 && - \left(\frac{12C_1}{r_{ij}^{14}}  - \frac{6C_2}{r_{ij}^{8}} \right) (I_3 - \mathbf{e}_{ij} \mathbf{e}_{ij}^{T}) \nonumber \\
 H_{vdW}(i,i) &=& - \sum_{j \in N(i)}  H_{col}(i,j) \nonumber \\
  \Delta(U_{vdW}) &=&    \sum_{i} \sum_{j \in N(i)} \left(\frac{132C_1}{r_{ij}^{14}}  - \frac{30C_2}{r_{ij}^{8} }\right) \nonumber \\
  \nabla\left( \Delta U_{vdW} \right) &=&    \sum_{i} \sum_{j \in N(i)} \left( -\frac{1848C_1}{r_{ij}^{15}}  +\frac{240C_2}{r_{ij}^{9}} \right) W_{ij}
 \end{eqnarray}
 
 The elastic potential is a simple quadratic potential:
  \begin{eqnarray}
 U_{el} &=& k_e \sum_{i} \sum_{j \in N(i)}  g(r_{ij}) r_{ij}^2,
 \end{eqnarray}
where $k_e = \frac{10\epsilon_G}{Npair}$, Npair is the number of pairs $(i,j)$, $N(i)$ is the cutoff-based neighborhood of $i$ already defined for the collision term, $g(r_{ij})$ is a Fermi-like function that defines a smooth cutoff:
\begin{eqnarray}
g(r) = \frac{1}{1 + \exp{\left(\frac{r-d_0}{a_0}\right)}}
\end{eqnarray}
where $d_0$ and $a_0$ are constants. We need three levels of derivatives for $g(r)$:
\begin{eqnarray} 
g'(x) &=& -\dfrac{1}{a_0} g(x)\big(1-g(x)\big) \nonumber \\
g''(x) &=& \dfrac{1}{a_0^{2}} g(x)\big(1-g(x)\big)\big(1-2g(x)\big) \nonumber \\
g'''(x) &=& -\dfrac{1}{a_0^{3}} g(x)\big(1-g(x)\big)\big(1-6g(x)+6g(x)^2\big) \nonumber \\
 \end{eqnarray}
 The gradient, Hessian, and Laplacian of the elastic potential are then derived as:
 \begin{eqnarray}
 \nabla U_{el} &=& k_e \sum_{i} \sum_{j \in N(i)} \left(g'(r_{ij}) r_{ij}^2 + 2g(r_{ij})r_{ij} \right) W_{ij} \nonumber \\
 H_{el}(i,j) &=& -k_e \left(g''(r_{ij})r_{ij}^2 +4g'(r_{ij})r_{ij} + 2 g(r_{ij}) \right) \mathbf{e}_{ij} \mathbf{e}_{ij}^{T}  \nonumber \\
 && - k_e \left(g'(r_{ij}) r_{ij} - 2g(r_{ij})\right) (I_3 - \mathbf{e}_{ij} \mathbf{e}_{ij}^{T}) \nonumber \\
 H_{el}(i,i) &=& - \sum_{j \in N(i)}  H_{el}(i,j) \nonumber \\
  \Delta(U_{el}) &=&   k_e  \sum_{i} \sum_{j \in N(i)} \left(2g''(r_{ij})r_{ij}^2 + 12g'(r_{ij}) + 12g(r_{ij})\right) \nonumber \\
  \nabla\left( \Delta U_{el} \right) &=&   k_e  \sum_{i} \sum_{j \in N(i)} \left( 2g'''(r_{ij})r_{ij}^2 + 16g''(r_{ij})r_{ij} \right. \nonumber \\
  && \quad \quad \quad \left.+ 24 g'(r_{ij}) \right) W_{ij} 
 \end{eqnarray}

%%%%%%%%%%%%%%%%%%%%%%%%%%%%%%%%%%%%%%%%%%%%%%%%%%%%%%%%%%%%
%%%%%%%%%%%%%%%%%%%%%%%%%%%%%%%%%%%%%%%%%%%%%%%%%%%%%%%%%%%%
\section{Angular potential: Gradient, Hessian, and Laplacian}
\label{sec:A2}
%%%%%%%%%%%%%%%%%%%%%%%%%%%%%%%%%%%%%%%%%%%%%%%%%%%%%%%%%%%%
%%%%%%%%%%%%%%%%%%%%%%%%%%%%%%%%%%%%%%%%%%%%%%%%%%%%%%%%%%%%

Let us consider the pseudo angle $\theta_i$ formed by three consecutive C$_\alpha$ along the backbone of a protein, $i$, $j$, and $k$, and centered at $j$. 
As above, we set the positions of these atoms as $\mathbf{r}_i$, $\mathbf{r}_j$, and $\mathbf{r}_k$. 
The role of the angular potential is to restrain this angle $\theta_j$ to match its value $\theta_j^0$ in the starting conformation of the protein:
\begin{eqnarray}
U_{\theta}(j) = K_a (\theta_j - \theta_j^0)^2.
\end{eqnarray}
Compared to the pairwise potentials defined in Appendix \ref{sec:A1}, $U_{\theta}(j)$ is a three-body potential. 
Computing its derivatives require care. 
We start with the simpler problem of computing the derivatives of $\theta_j$ with respect to the positions $r_i$, $r_j$, and $r_k$. 

%%%%%%%%%%%%%%%%%%%%%%%%%%%%%%%%%%%%%%%%%%%%%%%%%%%%%%%%%%%%
\subsection{Gradient, Hessian, and Laplacian of $\theta_j$:}
%%%%%%%%%%%%%%%%%%%%%%%%%%%%%%%%%%%%%%%%%%%%%%%%%%%%%%%%%%%%

Let $\mathbf{r}_{ij} = \mathbf{r_i} - \mathbf{r_j}$ and $\mathbf{r}_{kj} = \mathbf{r_k} - \mathbf{r_j}$.
Let $\mathbf{e}_{ij}$ and $\mathbf{e}_{kj}$ be the corresponding unit vectors.
We also set $r_{ij} =|\mathbf{r}_{ij}|$ and $r_{kj} =|\mathbf{r}_{kj}|$.
We introduce the cosine and sine of the angle at $j$:
\begin{eqnarray}
c = \mathbf{e}_{ij}\cdot\mathbf{e}_{kj}, \qquad s = \sqrt{1 - c^2}.
\end{eqnarray}.
The angle at vertex $j$ is then
\begin{eqnarray}
\theta_j = \arccos(c).
\end{eqnarray}

The cosine $c$ is a smooth function of $\mathbf{r}_{ij}$ and $\mathbf{r}_{kj}$:
\begin{eqnarray}
c = \frac{\mathbf{r}_{ij}\cdot\mathbf{r}_{kj}}{r_{ij} r_{kj}}
\end{eqnarray}.
Its first derivatives are
\begin{eqnarray}
\frac{\partial c}{\partial \mathbf{r}_{ij}} &= \frac{1}{r_{ij}} \left(\mathbf{e}_{kj} - c\mathbf{e}_{ij}\right), \nonumber \\
\frac{\partial c}{\partial \mathbf{r}_{kj}} &= \frac{1}{r_{kj}} \left(\mathbf{e}_{ij} - c\mathbf{e}_{kj}\right).
\end{eqnarray}

The second derivatives of $c$ with respect to $\mathbf{r}_{ij}$ and $\mathbf{r}_{kj}$  are $3\times3$ matrices:
\begin{eqnarray}
\frac{\partial^2 c}{\partial \mathbf{r}_{ij} \partial \mathbf{r}_{ij}}
  &=& \frac{1}{r_{ij}^2} 
 \left[-(\mathbf{e}_{kj}\mathbf{e}_{ij}^{T} + \mathbf{e}_{ij}\mathbf{e}_{kj}^{T})
      + 3c\mathbf{e}_{ij}\mathbf{e}_{ij}^{T} - c I_3\right], \nonumber \\
\frac{\partial^2 c}{\partial \mathbf{r}_{kj} \partial \mathbf{r}_{kj} }
  &=& \frac{1}{r_{kj}^2} \left[-(\mathbf{e}_{ij}\mathbf{e}_{kj}^{T} + \mathbf{e}_{kj}\mathbf{e}_{ij}^{T}) 
      + 3c\mathbf{e}_{kj}\mathbf{e}_{kj}^{T} - cI_3\right], \nonumber \\
\frac{\partial^2 c}{\mathbf{r}_{ij} \mathbf{r}_{kj}}
  &=& \frac{1}{r_{ij}r_{kj}} \left[I - \mathbf{e}_{ij}\mathbf{e}_{ij}^{T} - \mathbf{e}_{kj}\mathbf{e}_{kj}^{T} + c\mathbf{e}_{ij}\mathbf{e}_{kj}^{T}\right].
\end{eqnarray}

Since $\theta_j = \arccos(c)$, its derivatives follow from the chain rule:
\begin{eqnarray}
\frac{\partial \theta_j}{\mathbf{r}_{ij}} = -\frac{1}{s}\frac{\partial c}{\mathbf{r}_{ij}} \nonumber \\
\frac{\partial \theta_j}{\mathbf{r}_{kj}} = -\frac{1}{s}\frac{\partial c}{\mathbf{r}_{kj}}
\end{eqnarray}

Differentiating again gives the Hessian blocks for $\theta_j$ with respect to $\mathbf{r}_{ij}$ and $\mathbf{r}_{kj}$ (each \(3\times3\)):
\begin{eqnarray}
\frac{\partial^2 \theta_j}{\mathbf{r}_{ij} \mathbf{r}_{ij}}
  &=& -\frac{1}{s}\frac{\partial^2 c}{\mathbf{r}_{ij} \mathbf{r}_{ij}}
     -\frac{c}{s^3} \left(\frac{\partial c}{\mathbf{r}_{ij}}\frac{\partial c}{\mathbf{r}_{ij}}^{T}\right),\\
\frac{\partial^2 \theta_j}{\mathbf{r}_{kj} \mathbf{r}_{kj}}
  &=& -\frac{1}{s}\frac{\partial^2 c}{\mathbf{r}_{kj} \mathbf{r}_{kj}}
     -\frac{c}{s^3} \left(\frac{\partial c}{\mathbf{r}_{kj}}\frac{\partial c}{\mathbf{r}_{kj}}^{T}\right),\\
\frac{\partial^2 \theta_j}{\mathbf{r}_{ij} \mathbf{r}_{kj}}
  &=& -\frac{1}{s}\frac{\partial^2 c}{\mathbf{r}_{ij} \mathbf{r}_{kj}}
     -\frac{c}{s^3} \left(\frac{\partial c}{\mathbf{r}_{ij}}\frac{\partial c}{\mathbf{r}_{kj}}^{T}\right).
\end{eqnarray}

Using
\begin{eqnarray}
\frac{\partial \mathbf{r}_{ij} }{\partial \mathbf{r_i}_i}=I,\quad
\frac{\partial \mathbf{r}_{ij} }{\partial \mathbf{r_j}_i}=I,\quad
\frac{\partial \mathbf{r}_{kj} }{\partial \mathbf{r_k}_i}=I,\quad
\frac{\partial \mathbf{r}_{kj} }{\partial \mathbf{r_j}_i}=I,\quad
\end{eqnarray}

we can express the gradient and Hessian of $\theta_j$ with respect to the positions of $i$, $j$, and $k$:

\begin{eqnarray}
\nabla_{p_i}	\theta_j &=& \frac{\partial 	\theta}{\mathbf{r}_{ij}}, \nonumber \\
\nabla_{p_k}	\theta_j &=& \frac{\partial 	\theta}{\mathbf{r}_{kj}}, \nonumber \\
\nabla_{p_j}	\theta_j &=& -\frac{\partial 	\theta}{\mathbf{r}_{ij}} - \frac{\partial 	\theta}{\mathbf{r}_{kj}}.
\label{eqn:gradTheta}
\end{eqnarray}

The Hessian of $\theta_j$ can be written in block form:
\begin{eqnarray}
H_{\theta} =
\begin{bmatrix}
H_{ii} & H_{ij} & H_{ik} \\
H_{ji} & H_{jj} & H_{jk} \\
H_{ki} & H_{kj} & H_{kk}
\end{bmatrix},
\label{eqn:HTheta1}
\end{eqnarray}
where each block is a $3 \times 3$ matrix:

\begin{eqnarray}
H_{ii} &=& \frac{\partial^2\theta_j}{\mathbf{r}_{ij} \mathbf{r}_{ij}}, \nonumber \\
H_{ik} &=& \frac{\partial^2\theta_j}{\mathbf{r}_{ij} \mathbf{r}_{kj}}, \nonumber \\
H_{ij} &=& -\frac{\partial^2\theta_j}{\mathbf{r}_{ij} \mathbf{r}_{ij}} - \frac{\partial^2\theta}{\mathbf{r}_{ij} \mathbf{r}_{kj}}, \nonumber \\
H_{jk} &=& -\frac{\partial^2\theta_j}{\mathbf{r}_{kj} \mathbf{r}_{kj}} - \left(\frac{\partial^2\theta}{\mathbf{r}_{ij} \mathbf{r}_{kj}}\right)^{T}, \nonumber \\
H_{jj} &=& 
 \frac{\partial^2\theta_j}{\mathbf{r}_{ij} \mathbf{r}_{ij}}
+\frac{\partial^2\theta_j}{\mathbf{r}_{kj} \mathbf{r}_{kj}}
+\frac{\partial^2\theta_j}{\mathbf{r}_{ij} \mathbf{r}_{kj}}
+\left(\frac{\partial^2\theta}{\mathbf{r}_{ij}  \mathbf{r}_{kj}}\right)^{T}, \nonumber \\
H_{ki} &=& \left(H_{ik}\right)^{T}, \nonumber \\
H_{kj} &=& \left(H_{jk}\right)^{T}, \nonumber \\
H_{kk} &=& \frac{\partial^2\theta}{\mathbf{r}_{kj} \mathbf{r}_{kj}}.
\label{eqn:HTheta2}
\end{eqnarray}

Finally, we give a formula for the Laplacian of $\theta_j$, i.e., the trace of its Hessian:
\begin{eqnarray}
\Delta \theta_j = \operatorname{tr}(H)
   = \frac{2}{s} \left(\frac{c}{r_{ij}^{2}} - \frac{1}{r_{ij}r_{kj}} + \frac{c}{r_{kj}^{2}}\right).
\end{eqnarray}

Defining the auxiliary scalar
\begin{eqnarray*}
F = \frac{c}{r_{ij}^{2}} - \frac{1}{r_{ij}r_{kj}} + \frac{c}{r_{kj}^{2}},
\end{eqnarray*}
 the gradients of $\Delta \theta_j$ with respect to the edge vectors $\mathbf{r}_{ij}$ and $\mathbf{r}_{kj}$ are
\begin{eqnarray}
\frac{\partial \Delta \theta}{\partial \mathbf{r}_{ij}}
  &=& \frac{2}{s} 
      \left(-\frac{2c}{r_{ij}^{3}} + \frac{1}{r_{ij}^{2}r_{kj}}\right)\mathbf{e}_{ij} \nonumber \\
      &&+\frac{2}{s}  \left(\frac{1}{r_{ij}^{2}}+\frac{1}{r_{kj}^{2}}+\frac{Fc}{s^{2}}\right)\frac{\mathbf{e}_{kj}-c\mathbf{e}_{ij}}{r_{ij}}, \nonumber \\
\frac{\partial \Delta \theta}{\partial \mathbf{r}_{kj}}
  &=& \frac{2}{s} 
      \left(-\frac{2c}{r_{kj}^{3}} + \frac{1}{r_{ij}r_{kj}^{2}}\right)\mathbf{e}_{kj} \nonumber \\
      && +\frac{2}{s} \left(\frac{1}{r_{ij}^{2}}+\frac{1}{r_{kj}^{2}}+\frac{Fc}{s^{2}}\right)\frac{\mathbf{e}_{ij}-c\mathbf{e}_{kj}}{r_{kj}}.
\end{eqnarray}

Finally, the gradients of $\Delta \theta$ with respect to vertex positions are
\begin{eqnarray}
\nabla_{p_i}\Delta \theta &=& \frac{\partial \Delta \theta}{\partial  \mathbf{r}_{ij}}, \nonumber \\
\nabla_{p_k}\Delta \theta &=& \frac{\partial \Delta \theta}{\partial \mathbf{r}_{kj}}, \nonumber \\
\nabla_{p_j}\Delta \theta &=& -\frac{\partial \Delta \theta}{\partial  \mathbf{r}_{ij}} - \frac{\partial \Delta \theta}{\partial \mathbf{r}_{kj}}.
 \label{eqn:DTheta}
\end{eqnarray}

%%%%%%%%%%%%%%%%%%%%%%%%%%%%%%%%%%%%%%%%%%%%%%%%%%%%%%%%%%%%
\subsection{Gradient, Hessian, and Laplacian of $U_{\theta}(j)$:}
%%%%%%%%%%%%%%%%%%%%%%%%%%%%%%%%%%%%%%%%%%%%%%%%%%%%%%%%%%%%

We have everything we need to define the derivatives of $U_{\theta}(j)$. Fist, the gradient is given by:
\begin{eqnarray}
\nabla U_{\theta}(j) = 2 (\theta_j - \theta_j^0) \nabla \theta_j
\end{eqnarray}
where $\nabla \theta_j$ is given in equation \ref{eqn:gradTheta}.

The Hessian of $U_{\theta}(j)$ is given by:
\begin{eqnarray}
H(U_{\theta}(j)) = 2 (\nabla \theta_j) (\nabla \theta_j)^T + 2 (\theta_j - \theta_j^0) H_{\theta}
\end{eqnarray}
with $H_{\theta}$ defined in equations \ref{eqn:HTheta1} and \ref{eqn:HTheta2}.

The Laplacian of $U_{\theta}(j)$ is equal to:
\begin{eqnarray}
\Delta U_{\theta}(j) &=& 2 (\nabla \theta_j)^2 + 2 (\theta_j - \theta_j^0) \Delta \theta_j
\end{eqnarray}
where $\Delta \theta_j$ is given in equation \ref{eqn:DTheta}.

Finally, the derivatives of the Laplacian of $U_{\theta}(j)j$ are given by:
\begin{eqnarray}
\nabla \Delta U_{\theta}(j) &=& 4  H_{\theta} \nabla \theta  + 2  \Delta \theta_j \nabla \theta + \nonumber \\
&& \quad \quad 2 (\theta_j - \theta_j^0) \nabla \Delta \theta
\end{eqnarray}

%%%%%%%%%%%%%%%%%%%%%%%%%%%%%%%%%%%%%%%%%%%%%%%%%%%%%%%%%%%%
%%%%%%%%%%%%%%%%%%%%%%%%%%%%%%%%%%%%%%%%%%%%%%%%%%%%%%%%%%%%
\section{The effective potential}
\label{sec:A3}
%%%%%%%%%%%%%%%%%%%%%%%%%%%%%%%%%%%%%%%%%%%%%%%%%%%%%%%%%%%%
%%%%%%%%%%%%%%%%%%%%%%%%%%%%%%%%%%%%%%%%%%%%%%%%%%%%%%%%%%%%

Recall that the potential $W$ is defined by:
\begin{eqnarray}
W = \frac{1}{4} \left (\nabla U \right)^2 - \frac{kT}{2} \Delta U
\end{eqnarray}
where $U$ is the full potential, namely the sum of the bond, angle, collision, and elastic potentials.
While $U$ is linear in those potentials, it is not the case of $W$, because of the term $\left(\nabla U \right)$.
However, we are really interested in the gradient of $W$, given by:
\begin{eqnarray}
\nabla W = \frac{1}{2} H_U \nabla U - \frac{kT}{2} \nabla \Delta U
\end{eqnarray}
We have, by linearity,
\begin{eqnarray}
\nabla U &=& \nabla U_b +  \nabla U_{\theta} + \nabla U_{vdW} + \nabla U_{el} \nonumber \\
H_U &=&  H_b +  H_{ang} + H_{col} + H_{el} \nonumber \\
\Delta E &=& \Delta U_b +  \Delta U_{\theta} + \Delta U_{vdW} + \Delta U_{el},
\end{eqnarray}
with all these terms defined in appendices \ref{sec:A1} and \ref{sec:A2}.

%%%%%%%%%%%%%%%%%%%%%%%%%%%%%%%%%%%%%%%%%%%%%%%%%%%%%%%%%%%%
%%%%%%%%%%%%%%%%%%%%%%%%%%%%%%%%%%%%%%%%%%%%%%%%%%%%%%%%%%%%

%\nocite{*}
\bibliography{langevin}% Produces the bibliography via BibTeX.

%%%%%%%%%%%%%%%%%%%%%%%%%%%%%%%%%%%%%%%%%%%%%%%%%%%%%%%%%%%%
%%%%%%%%%%%%%%%%%%%%%%%%%%%%%%%%%%%%%%%%%%%%%%%%%%%%%%%%%%%%

\end{document}